# Approximately achieving Gaussian relay network capacity with lattice-based QMF codes

Ayfer Özgür and Suhas Diggavi

**Abstract**

Recently, it has been shown that a new relaying strategy, quantize-map-and-forward (QMF) scheme approximately achieves (within an additive constant number of bits) the Gaussian relay network capacity for arbitrary topologies [1]. This was established using Gaussian codebooks for transmission and random mappings at the relays. In this paper, we show that a similar approximation result can be established by using lattices for transmission and quantization along with structured lattice-to-lattice mappings at the relays. We establish this result for both full duplex and half duplex wireless relay networks.

## I. INTRODUCTION

Characterizing the capacity of relay networks has been a long-standing open question in network information theory. The seminal work of Cover and El-Gamal [6] has established several basic achievability schemes for relay channels. More recently there has been extension of these techniques to larger networks (see [9] and references therein). In [1], motivated by a deterministic model of wireless communication, a new relaying strategy, called quantize-map-forward (QMF) was developed. It was shown that the quantize-map-and-forward scheme achieves within a constant number of bits from the information-theoretic cutset upper bound. This constant is universal in the sense that it is independent of the channel gains and the operating SNR, though it could depend on the network topology (like the number of nodes). Moreover QMF was shown to be robust in that the relays did not need information about network topology or channel conditions, and it also achieved the compound network capacity approximately.

In the QMF scheme developed in [1], each relay node first quantizes its received signal at the noise level, then randomly maps it directly to a Gaussian codeword and transmits it. Following this result, there have been several papers that build on this approximation strategy (see for example [2], [5], [3], [7] and references therein). A natural question that we address in this paper is whether lattice codes retain the approximate optimality of the above scheme. This is motivated in part since lattice codes along with lattice decoding could enable computationally tractable encoding and decoding methods. For example lattice codes were used to achieve the capacity of Gaussian channels in [15], and for communication over multiple-access relay networks (with orthogonal broadcast) in [14].

The main result of this paper is to show that the QMF scheme using nested lattice codes for transmission and quantization along with structured lattice-to-lattice maps, still achieves the Gaussian relay network capacity within a constant. This result[1], is summarized in Theorem 2.1. It also enables many other approximation results established in [1]; all those approximation results can now be achieved through structured lattice codes. These include the result for multicast networks as well as for compound networks.

The use of structured lattice codes requires the specification of a structured lattice-to-lattice mapping between the quantization and transmission codebooks at each relay. We design such a map by using the representation of nested lattices through linear codes lifted appropriately to the real domain. Such a representation of nested lattices was studied in [18] and also [15]. This enables us to design lattice-to-lattice maps at the relays that are implementable with polynomial complexity, but still retain approximate optimality. In this paper we make several other technical contributions to establish the main result: (i) we use a lattice vector quantizer instead of the scalar lattice quantizer used in [1], and this enables us to get a better approximation constant. (ii) we develop a "typical decoder" analysis for lattices that enables us to establish the approximation result, which might be of independent interest. (iii) we develop a simple outer bound on the information-theoretic capacity of half-duplex networks, earlier upper bounds apply under the restriction of fixed schedules and no transmit power optimization across the half-duplex states as explained below.

Half-duplex radios have the constraint that they cannot transmit and receive signals simultaneously over the same frequency band. Therefore, each relay needs to develop a strategy of when to listen and when to transmit. Fixed scheduling strategies are those where the listen-talk states of the relays are established prior to the start of communication (but perhaps depending on global channel/network conditions). However, random scheduling strategies are those which allow the schedules to change during run-time, so that the transmit and receive states of the relays can be used to convey additional information. Moreover, the transmit power of the relays can be optimized across different configurations of the network. Note that in a network of $N$ relays where each relay can be in either transmit or receive state, there are $2^N$ different possible configurations for the network. We show that the QMF strategy with fixed schedules and an equal power allocation strategy across the half-duplex states, can approximately achieve the capacity of half-duplex networks. This establishes the first approximation result for half-duplex networks. Note that earlier approximation results were based on restricting to fixed scheduling strategies with equal power allocation [1]. It is easy to observe that the random strategies can increase the capacity by at most one bit per relay over fixed schedules, or $N$ bits/s/Hz in total. This has been pointed out in [12], [13]. However, to the best of our knowledge, the capacity gain due to transmit power optimization across the $2^N$ states of the network has not been investigated earlier. We show that this gain can be at most linear in $N$.

The paper is organized as follows: In Section II, we state the network model and our main results. In Section III, we summarize the construction of the nested lattice ensemble. In Section IV, we describe the network operation. In particular, we specify how we use the nested lattice codes of Section III for encoding at the source, quantization, lattice-to-lattice mapping and transmission at the relay nodes, and decoding at the destination node. In Section V, we analyze the performance achieved by the scheme. In Section VI, we establish the approximation result for half-duplex networks. Many of the detailed proofs are given in the Appendices.

---

[1]This result was first presented in [4], is the first structured code for approximately achieving the wireless network capacity.

## II. MAIN RESULTS

We consider a Gaussian relay network with a set $\mathcal{M}$ of $N$ nodes, where a source node $s \in \mathcal{M}$ wants to communicate to a destination node $d \in \mathcal{M}$, with the help of relay nodes $\mathcal{M} \setminus \{s, d\}$. The signal received by node $i \in \mathcal{M}$ is given by

$$\mathbf{y}_i = \sum_{j \neq i} H_{ij} \mathbf{x}_j + \mathbf{z}_i \tag{1}$$

where $H_{ij}$ is the $N_i \times M_j$ channel matrix from node $j$ comprising $M_j$ transmit antennas to node $i$ comprising $N_i$ receive antennas. Each element of $H_{ij}$ represents the complex channel gain from a transmitting antenna of node $j$ to a receiving antenna of node $i$. The noise $\mathbf{z}_i$ is complex circularly-symmetric Gaussian vector $\mathcal{CN}(0, I)$ and is i.i.d. for different nodes. The transmitted signals $\mathbf{x}_j$ are subject to an average power constraint $P$. Note that without loss of generality we have scaled the noise power to $1$.

The following theorems are the main result of this paper.

*Theorem 2.1:* Using nested lattice codes for transmission and quantization along with structured mappings at the relays, we can achieve all rates

$$R \leq \min_{\Omega} I(\mathbf{x}_\Omega; \mathbf{y}_{\Omega^c} | \mathbf{x}_{\Omega^c}) - (2 + \log 2) \sum_{i \in \mathcal{M} \setminus s} N_i$$

between $s$ and $d$, where $\Omega$ is a source-destination cut of the network, $\mathbf{x}_\Omega = \{\mathbf{x}_i, i \in \Omega\}$ and $\mathbf{x}_i, i \in \mathcal{M}$ are i.i.d. $\mathcal{CN}(0, (P/M_i)I)$.[2]

It has been shown in [1] (see Lemma 6.6) that the restriction to i.i.d. Gaussian input distributions is within $2 \sum_{i \in \mathcal{M}} M_i$ bits/s/Hz of the cut-set upper bound. Therefore the rate achieved using lattice codes in the above theorem is within $2 \sum_{i \in \mathcal{M}} M_i + (2 + \log 2) \sum_{i \in \mathcal{M}} N_i$ bits/s/Hz to the cutset upper bound of the network (or $\sum_{i \in \mathcal{M}} M_i + (2 + \log 2)/2 \sum_{i \in \mathcal{M}} N_i$ for real Gaussian networks)[3]. This is summarized in the following result.

*Theorem 2.2:* Using nested lattice codes, we can approximately achieve the capacity of Gaussian wireless networks to within $2 \sum_{i \in \mathcal{M}} M_i + (2 + \log 2) \sum_{i \in \mathcal{M}} N_i$ bits/s/Hz.

The same lattice coding techniques used to obtain the approximate characterization of Theorem 2.2, can be used to get the approximate characterization for multiple-source multicast (where there are multiple sources and destinations, which are interested in all the sources) as well as for compound relay networks. The extensions of lattice codes to these cases are straightforward applications of the ideas in this paper using the tools developed in [1], [10], [11]. Another interesting case is that of half-duplex networks considered in [1], where it was established that the QMF scheme approximately achieved the best possible rates, when attention was restricted to the class of fixed schedules with constant transmit power for the relays. In this paper, we establish the approximation result for *any* scheme over half duplex networks. Moreover, we show that a uniform power allocation across the states is approximately optimal. The following result is proved in Section VI.

---

[2] The logarithms in the paper are base $e$.

[3] These constants can be further tightened by using a sharper analysis and adjusting the quantization levels, but our goal here is not to get the tightest bound for the constants.

*Theorem 2.3:* Using nested lattice codes and fixed scheduling of transmission states, we can approximately achieve the capacity of Gaussian relay networks with half-duplex constraint to within $N + 4\sum_{i\in\mathcal{M}} M_i + (2 + \log 2)\sum_{i\in\mathcal{M}} N_i$ bits/s/Hz.

For simplicity of presentation, in the rest of the paper we concentrate on scalar channels where every node has a single transmit and receive antenna. Moreover, we focus our attention to *layered networks*, which were defined in [1]. These are networks, where the number of hops are the same for every path from the source to the destination in the network. An example of such a layered network is given in Figure 1. More precisely, the signal received by node $i$ in layer $l, 0 \leq l \leq l_d$, denoted $i \in \mathcal{M}_l$, is given by

$$\mathbf{y}_i = \sum_{j\in\mathcal{M}_{l-1}} h_{ij}\mathbf{x}_j + \mathbf{z}_i$$

where $h_{ij}$ is the real scalar channel coefficient from node $j$ to node $i$ and $s \in \mathcal{M}_0, d \in \mathcal{M}_{l_d}$. The analysis can be extended to arbitrary (non-layered) networks by following the time-expansion argument of [1] and to multicast traffic with multiple destination nodes as well as to multiple multicast where multiple source nodes multicast to a group of destination nodes. The complex case follows by representing each complex number as a two-dimensional real vector. The extension to multiple antennas is discussed inside the text.

## III. Preliminaries: Construction of the Nested Lattice Ensemble

In this section we review some of the basic properties of lattices that can be found in standard references like [15], [16], [17]. We summarize these properties to make this paper more self-contained, as well as to establish the notation used throughout this paper.

Consider a lattice $\Lambda$, or more precisely, a sequence of lattices $\Lambda^{(n)}$ indexed by the lattice dimension $n$, with $\mathcal{V}$ denoting the Voronoi region of $\Lambda$. The second moment per dimension of $\Lambda$ is defined as

$$\sigma^2(\Lambda) = \frac{1}{n}\frac{1}{|\mathcal{V}|}\int_{\mathcal{V}} \|\mathbf{x}\|^2 d\mathbf{x}$$

where $|\mathcal{V}|$ denotes the volume of $\mathcal{V}$. We also define the normalized second moment of $\Lambda$,

$$G(\Lambda) = \frac{\sigma^2(\Lambda)}{|\mathcal{V}|^{2/n}}. \qquad (2)$$

Throughout the paper, we assume that $\Lambda$ (or more precisely, the sequence of lattices $\Lambda^{(n)}$) is both Rogers and Poltyrev-good. The existence of such lattices has been shown in [16]. Formally, $\Lambda$ satisfies the following properties:

- (Rogers-good) Let $R_u$ and $R_l$ be the covering and effective radius of the lattice $\Lambda$. $\Lambda$ (more precisely the sequence of lattices $\Lambda^{(n)}$) is called Rogers-good if its covering efficiency approaches 1 as the dimension $n$ grows,

$$\rho_{cov}(\Lambda) = \frac{R_u}{R_l} \to 1. \qquad (3)$$

It is known that a lattice that is good for covering is necessarily good for quantization. A lattice is called good for quantization if

$$G(\Lambda) \to G_n^* \tag{4}$$

where $G_n^*$ is the normalized second moment of an $n$-dimensional sphere and $G_n^* \to \frac{1}{2\pi e}$ when the dimension $n$ becomes large. (4) follows from (3) and the relation (see [16])

$$G(\Lambda) \leq \frac{n+2}{n} G_n^* (\rho_{cov})^2.$$

- (Poltyrev-good) Let $Z$ be a Gaussian random vector whose components are i.i.d. $\mathcal{N}(0, \sigma^2)$, such that $\sigma^2 \leq \sigma^2(\Lambda)$. The volume to noise ratio of the lattice $\Lambda$ relative to $\mathcal{N}(0, \sigma^2)$ is defined as $\mu = \sigma^2(\Lambda)/\sigma^2$. Then, $\Lambda$ (more precisely the sequence of such lattices $\Lambda^{(n)}$) is called Poltyrev-good if

$$\mathbb{P}(\mathbf{Z} \notin \mathcal{V}) < e^{-n[E_P(\mu) - o_n(1)]}$$

where $E_P(\mu)$ is the Poltyrev exponent given by

$$E_P(\mu_i) = \begin{cases} \frac{1}{2}[(\mu_i - 1) - \log \mu_i] & 1 < \mu_i \leq 2 \\ \frac{1}{2} \log \frac{e\mu_i}{4} & 2 \leq \mu_i \leq 4 \\ \frac{\mu_i}{8} & \mu_i \geq 4. \end{cases}$$

Let the $n \times n$ full-rank generator matrix of $\Lambda$ be denoted by $G_\Lambda$, i.e., $\Lambda = G_\Lambda \mathbb{Z}^n$.[4] This fixed lattice $\Lambda$ will serve as the coarse lattice for all the nested lattice constructions in this paper. The fine lattice $\Lambda_1$ is constructed using Loeliger's type-A construction [18]. Let $m, n, p$ be integers such that $m \leq n$ and $p$ is prime. The fine lattice is constructed using the following steps:

- Draw an $n \times m$ matrix $G$ such that each of its entries is i.i.d according to the uniform distribution over $\mathbb{Z}_p = \{0, 1, \ldots, p-1\}$.
- Form the linear code

$$\mathcal{C} = \{\mathbf{c} : \mathbf{c} = G \cdot \mathbf{w}, \mathbf{w} \in \mathbb{Z}_p^m\}, \tag{5}$$

where "$\cdot$" denotes modulo-p multiplication.
- Lift $\mathcal{C}$ to $\mathbb{R}^n$ to form [5]

$$\Lambda_1' = p^{-1}\mathcal{C} + \mathbb{Z}^n.$$

where for two sets $A \subset \mathbb{R}^n$ and $B \subset \mathbb{R}^n$, the sum set $A + B \subset \mathbb{R}^n$ denotes $A + B = \{\mathbf{a} + \mathbf{b} : \mathbf{a} \in A, \mathbf{b} \in B\}$.
- $\Lambda_1 = G_\Lambda \Lambda_1'$ is the desired fine lattice. Note that since $\mathbb{Z}^n \subseteq \Lambda_1'$, we have $\Lambda \subseteq \Lambda_1$.

---

[4]For any operation $f : \mathbb{R}^n \to \mathbb{R}^n$ and a set $A \subset \mathbb{R}^n$, $f(A) \subset \mathbb{R}^n$ denotes $f(A) = \{f(a) : \mathbf{a} \in A\}$.

[5]In the sequel, we slightly abuse notation by using $\mathcal{C}$ to denote both the code over the finite field and its projection to the reals. Hence, the codewords $c$ are either considered as vectors in $\mathbb{Z}_p^n$, in which case they are subject to finite field operations, or they are considered as vectors in $\mathbb{R}^n$ subject to real field operations. It is to be deduced from the context to which of these two cases the notation refers to.

- Draw $\mathbf{v}$ uniformly over $p^{-1}\Lambda \cap \mathcal{V}$ and translate the lattice $\Lambda_1$ by $\mathbf{v}$. The nested lattice codebook consists of all points of the translated fine lattice inside the Voronoi region of the coarse lattice,

$$\Lambda^* = (\mathbf{v} + \Lambda_1) \mod \Lambda = (\mathbf{v} + \Lambda_1) \cap \mathcal{V}. \tag{6}$$

In the above equation, we define $\mathbf{x} \mod \Lambda$ as the quantization error of $\mathbf{x} \in \mathbb{R}^n$ with respect to the lattice $\Lambda$, i.e.,

$$\mathbf{x} \mod \Lambda = \mathbf{x} - Q_\Lambda(\mathbf{x}), \tag{7}$$

where $Q_\Lambda(\mathbf{x}) : \mathbb{R}^n \to \Lambda$ is the nearest-neighbor lattice quantizer defined as,

$$Q_\Lambda(\mathbf{x}) = \arg\min_{\lambda \in \Lambda} \|\mathbf{x} - \lambda\|.$$

Note that the quantization and mod operations with respect to a lattice can be defined in different ways. The mod operation in (7) maps $\mathbf{x} \in \mathbb{R}^n$ to the Voronoi region $\mathcal{V}$ of the lattice. More generally, it is possible to define a mod or quantization operation with respect to any fundamental region of the lattice. In particular, when we consider the integer lattice $\mathbb{Z}^n$ in the sequel, or more generally its multiples $p\mathbb{Z}^n$ where $p$ is a positive integer, we will assume that

$$\mathbf{x} \mod p\mathbb{Z}^n = \mathbf{x} - \lfloor \mathbf{x} \rfloor_p$$

where $\lfloor \mathbf{x} \rfloor_p$ denotes component-wise rounding to the nearest smaller integer multiple of $p$. In other words, the mod operation with respect to $p\mathbb{Z}^n$ will map the point $\mathbf{x} \in \mathbb{R}^n$ to the region $p[0,1)^n$.

The above construction yields a random ensemble of nested lattice codes that has a number of desired properties as we discuss next.

First, note that there is a bijection between

$$\mathbb{Z}_p^n \leftrightarrow p^{-1}\mathbb{Z}_p^n = p^{-1}\mathbb{Z}^n \cap [0,1)^n \leftrightarrow p^{-1}\Lambda \cap G_\Lambda [0,1)^n \leftrightarrow p^{-1}\Lambda \cap \mathcal{V}.$$

The last bijection follows from the fact that both $G_\Lambda [0,1)^n$ and $\mathcal{V}$ are fundamental regions of the lattice $\Lambda$, i.e., they both tile $\mathbb{R}^n$. Since $\mathcal{C} \subseteq \mathbb{Z}_p^n$, the above bijection restricted to $\mathcal{C}$ yields,

$$\mathcal{C} \leftrightarrow p^{-1}\mathcal{C} = \Lambda_1' \cap [0,1)^n \leftrightarrow \Lambda_1 \cap G_\Lambda [0,1)^n \leftrightarrow \Lambda_1 \cap \mathcal{V} \leftrightarrow \Lambda^*. \tag{8}$$

Note also that $\Lambda^* \subseteq p^{-1}\Lambda \cap \mathcal{V}$. The bijections above can be explicitly specified in both directions and we will make use of this fact in the next section.

Note that $\mathbf{w}$ in (5) runs through all the $p^m$ vectors in $\mathbb{Z}_p^m$. Let us index these vectors as $\mathbf{w}(i)$, $i = 0, \ldots, p^m - 1$. Let us index the corresponding codewords in $\mathcal{C}$ as $\mathcal{C}(i) = G \cdot \mathbf{w}(i)$, $i = 0, \ldots, p^m - 1$. The $p^m$ codewords in $\mathcal{C}$ need not be distinct. By the bijection in (8), each codeword in $\mathcal{C}$ corresponds to one fine lattice point in $\Lambda_1 \cap \mathcal{V}$ and one codeword of $\Lambda^*$. Let us similarly index the points in $\Lambda_1 \cap \mathcal{V}$ as $\Lambda_1(i)$ and the corresponding codewords of $\Lambda^*$ as $\Lambda^*(i)$, for $i = 0, \ldots, p^m - 1$. We have,

$$\Lambda_1(i) = G_\Lambda p^{-1}\mathcal{C}(i) \mod \Lambda \qquad \Lambda^*(i) = (\mathbf{v} + \Lambda_1(i)) \mod \Lambda. \tag{9}$$

*Proposition 3.1:* The random codebook $\Lambda^*$ defined in (9) has the following statistical properties:

- Let $\lambda \in p^{-1}\Lambda \cap \mathcal{V}$,
$$\mathbb{P}(\Lambda^*(i) = \lambda) = \frac{1}{|p^{-1}\Lambda \cap \mathcal{V}|} = \frac{1}{p^n}. \tag{10}$$

- Let $\lambda_1, \lambda_2 \in p^{-1}\Lambda \cap \mathcal{V}, \quad \forall i \neq j$,
$$\mathbb{P}(\Lambda^*(i) = \lambda_1, \Lambda^*(j) = \lambda_2) = \frac{1}{|p^{-1}\Lambda \cap \mathcal{V}|^2} = \frac{1}{p^{2n}}. \tag{11}$$

In other words, the construction in this section yields an ensemble of nested lattice codes such that each codeword of the random codebook $\Lambda^*$ is uniformly distributed over $p^{-1}\Lambda \cap \mathcal{V}$ and the codewords of $\Lambda^*$ are pairwise independent. These two properties suffice to prove the random coding result of this paper.

*Proof of Proposition 3.1* The first property (10) simply follows from the fact that $\mathbf{v}$ is uniformly distributed on $p^{-1}\Lambda \cap \mathcal{V}$. For the second probability, we have

$$\begin{aligned}
&\mathbb{P}(\Lambda^*(i) = \lambda_1, \Lambda^*(j) = \lambda_2) \\
&= \mathbb{P}\left((\mathbf{v} + \Lambda_1(i)) \mod \Lambda = \lambda_1, \ (\mathbf{v} + \Lambda_1(j)) \mod \Lambda = \lambda_2\right) \\
&= \mathbb{P}((\mathbf{v} + \Lambda_1(i)) \mod \Lambda = \lambda_1, \ (\mathbf{v} + \Lambda_1(j)) \mod \Lambda - (\mathbf{v} + \Lambda_1(i)) \mod \Lambda = \lambda_2 - \lambda_1) \\
&= \mathbb{P}(\Lambda_1(i) = (\lambda_1 - \mathbf{v}) \mod \Lambda, \ (\Lambda_1(j) - \Lambda_1(i)) \mod \Lambda = (\lambda_2 - \lambda_1) \mod \Lambda) \\
&= \mathbb{P}((\Lambda_1(j) - \Lambda_1(i)) \mod \Lambda = (\lambda_2 - \lambda_1) \mod \Lambda) \\
&\quad \times \mathbb{P}(\Lambda_1(i) = (\lambda_1 - \mathbf{v}) \mod \Lambda \mid (\Lambda_1(j) - \Lambda_1(i)) \mod \Lambda = (\lambda_2 - \lambda_1) \mod \Lambda). \tag{12}
\end{aligned}$$

Note that the first probability in (12) is independent of $\mathbf{v}$. Let us denote $\lambda = (\lambda_2 - \lambda_1) \mod \Lambda \in p^{-1}\Lambda \cap \mathcal{V}$, we have

$$\begin{aligned}
(\Lambda_1(j) - \Lambda_1(i)) \mod \Lambda = \lambda &\Leftrightarrow (G_\Lambda p^{-1}\mathcal{C}(j) \mod \Lambda - G_\Lambda p^{-1}\mathcal{C}(i) \mod \Lambda) \mod \Lambda = \lambda \\
&\Leftrightarrow (G_\Lambda p^{-1}\mathcal{C}(j) - G_\Lambda p^{-1}\mathcal{C}(i)) \mod \Lambda = \lambda \\
&\Leftrightarrow (G_\Lambda p^{-1}\mathcal{C}(j) - G_\Lambda p^{-1}\mathcal{C}(i)) = \lambda + \mathbf{x}, \quad \mathbf{x} \in \Lambda \\
&\Leftrightarrow (\mathcal{C}(j) - \mathcal{C}(i)) = p G_\Lambda^{-1}\lambda + p G_\Lambda^{-1}\mathbf{x}, \quad p G_\Lambda^{-1}\mathbf{x} \in p\mathbb{Z}^n \\
&\Leftrightarrow (\mathcal{C}(j) - \mathcal{C}(i)) \mod p\mathbb{Z}^n = p G_\Lambda^{-1}\lambda \mod p\mathbb{Z}^n \tag{13} \\
&\Leftrightarrow G \cdot (\mathbf{w}(j) - \mathbf{w}(i)) = \mathbf{c}, \tag{14}
\end{aligned}$$

where all equations except the last one are over the reals. The last equation (14) is a restatement of (13) in terms of finite field operations with $\mathbf{c} = p G_\Lambda^{-1}\lambda \mod p\mathbb{Z}^n$ in (14) treated as a finite-field vector in $\mathbb{Z}_p^n$. Since $j \neq i$, the vector $\mathbf{w}(j) - \mathbf{w}(i)$ has at least one nonzero entry. Since the corresponding column of $G$ is uniformly distributed over $\mathbb{Z}_p^n$, we have

$$\mathbb{P}(G \cdot (\mathbf{w}(j) - \mathbf{w}(i)) = \mathbf{c}) = \mathbb{P}((\Lambda_1(j) - \Lambda_1(i)) \mod \Lambda = (\lambda_2 - \lambda_1) \mod \Lambda) = \frac{1}{p^n}.$$

For the second probability in (12), it is easy to observe that for any realization of $G$, hence $\Lambda_1(i)$, there is exactly one choice of $\mathbf{v}$ out of $p^n$ possible choices that satisfies the equality $\Lambda_1(i) = (\lambda_1 - \mathbf{v}) \mod \Lambda$. Combining these observations yields the conclusion in (11). $\square$

The above construction yields a random ensemble of nested lattice pairs $\Lambda \subseteq \Lambda_1$ with coding rate,

$$R = \frac{1}{n} \log |\Lambda^*|$$

which can be tuned by choosing the precise magnitudes of $m$ and $p$. Note that $|\Lambda^*| = p^m$ if the random matrix $G$ in (5) is full rank. The probability that $G$ is not full rank can be upper bounded by

$$\mathbb{P}(\mathrm{rank}(G) < m) = \sum_{\mathbf{w} \in \mathbb{Z}_p^m, \mathbf{w} \neq \mathbf{0}} \mathbb{P}(G \cdot \mathbf{w} = \mathbf{0}) = (p^m - 1) \, p^{-n}.$$

Therefore if $m \leq \beta n$ for $\beta < 1$, the above probability decreases to zero at least exponentially as $n$ increases ($p$ may also grow with $n$). We assume that $m$ is chosen to satisfy this condition in all our nested lattice code constructions in the next section.

## IV. Lattice based QMF scheme

The quantize-map-forward (QMF) strategy, introduced in [1] is the following. Each relay first quantizes the received signal at the noise level, then randomly maps it to a Gaussian codeword and transmits it. The destination then decodes the transmitted message, without requiring the decoding of the quantized values at the relays. This overall operation ensures that the relays need not know the network topology, or the channel gains of the signals being received by it[6]. The specific scheme that [1] focused on was based on a scalar (lattice) quantizer followed by a mapping to a Gaussian random codebook. However, the use of vector quantizers and Gaussian codebooks leads to similar approximation results (see [2], [11] and references therein). However, the focus of this paper is to use lattices in order to implement the QMF scheme and analyze it.

We first replace the (Gaussian) quantizer and the Gaussian transmit codebook at each relay with lattice versions. This basically leads us to design lattice-to-lattice maps at the relays. Intuitively, this is done by using the linear code representation of the lattices described in Section III. Once the relay quantizes the received signal, using the bijection given in (8) we can extract the point $\mathbf{c}$ in the finite field corresponding to the quantized value $\hat{\mathbf{y}}$. Now, this point is linearly transformed using a random matrix $G$ over the finite field, and then $G\mathbf{c}$ is viewed as a finite field representation of the transmit lattice $\Lambda$. Therefore it can be "lifted" to the real domain and transmitted. This intuition is made precise in (22) and Proposition 4.2. Note that this transformation effectively only requires a matrix multiplication over the finite field and hence has *polynomial complexity* in the number of operations required to implement it[7].

As mentioned earlier, description of the lattice-based scheme and its analysis (in Section V) will be done for layered networks (illustrated in Figure 1). However, the extension of these results to arbitrary (non-layered) networks is done through the standard technique of time-expansion (see [1], Section VI B). In order to implement the QMF scheme, we also need to specify the decoder used by the destination. For this, we define a lattice-based

---

[6]Of course the final destination, which needs to decode the source message needs to know these channels to be able to unravel the transformations to decode.

[7]This is assuming that the quantization to the lattice point can be done efficiently. This is true for integer lattices.

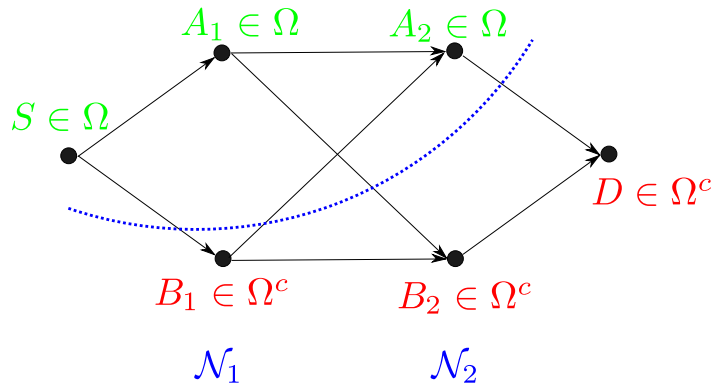

Fig. 1. Example of a layered network, where all paths from $S$ to $D$ are three hops. Additionally this clarifies the notation $\mathcal{M}_l$ for the $l^{th}$ layer, where $\mathcal{M}_1 = \{A_1, B_1\}$ and $\mathcal{M}_1 = \{A_2, B_2\}$, with $l_d = 3$, implying that $\mathcal{M}_{L_d} = \{D\}$.

"typicality" decoder[8]. Such a decoder finds a "plausible" sequence of received (quantized) sequences that could have resulted in the received observation. Given this definition, we can bound the probabilities using appropriate Gaussian approximation and therefore use an analysis inspired by [1]. A more precise definition of the lattice typicality decoder is given in (26) and the precise analysis is done in Section V.

In the previous section, we have constructed an ensemble of nested lattices where the coarse lattice $\Lambda$ is fixed and the fine lattice $\Lambda_1$ is randomized. It has been shown in [19] that with high probability, a nested lattice $(\Lambda_1, \Lambda)$ in this ensemble is such that both $\Lambda_1$ and $\Lambda$ are Rogers and Poltyrev-good. (The fixed lattice $\Lambda$ is Rogers and Poltyrev-good by construction.) For quantization and transmission at each relay, we use randomly and independently generated codebooks by the construction of the earlier section. Even though we use the same construction, the codebooks are generated with different parameters depending on whether we do transmission or quantization and also depending on the noise level at each relay. The mapping between the quantization and transmission codebooks at each relay is specified below.

**Source:** The source has $p_s{}^{m_s}$ messages, where $p_s$ is prime and $m_s \leq n$. The messages are represented as length-$m_s$ vectors over the finite field $\mathbb{Z}_{p_s}$ and mapped to a random nested lattice codebook $\Lambda^*$ following the construction in Section III. In the construction, the coarse lattice $\Lambda$ is scaled such that its second moment,

$$\sigma^2(\Lambda^T) = \frac{n}{n+2} \frac{G(\Lambda^T)}{G_n^*} \frac{1}{(\rho_{cov}(\Lambda^T))^2} P, \tag{15}$$

where $\Lambda^T$ now denotes the scaled version of the lattice $\Lambda$ to satisfy the power constraint. Note that $\sigma^2(\Lambda^T) \to P$ as $n$ increases since $\Lambda^T$ is Rogers-good. This choice ensures that every codeword of $\Lambda^*$ satisfies the power constraint $P$. This result is stated in the Proposition 4.1 below. The information rate of the code is given by

$$R = \frac{1}{n} \log p_s{}^{m_s}.$$

---

[8]The definition of the typicality decoder for lattices is inspired by the Gaussian version. This might be independently useful for any lattice based scheme.

Let us denote by $\mathbf{x}_s^{(w)}$, $w \in \{1, \ldots, e^{nR}\}$ the random transmit codewords corresponding to each message $w$ of the source node. Note that by Proposition 3.1, the messages $w$ are mapped uniformly and pair-wise independently to the lattice points $p^{-1}\Lambda^T \cap \mathcal{V}^T$.

*Proposition 4.1:* Each transmitted codeword $\mathbf{x}_s^{(w)}$ satisfies the transmit power constraint $P$.

*Proof of Proposition 4.1:* Since every transmitted codeword $\mathbf{x}_s^{(w)} \in \mathcal{V}^T$, we have

$$\frac{1}{n}\|\mathbf{x}_s^{(w)}\|^2 \leq \frac{1}{n}(R_u^T)^2,$$

where $R_u^T$ is the covering radius of $\Lambda^T$. We now relate the covering radius $R_u^T$ of $\Lambda^T$ to its second moment $\sigma^2(\Lambda^T)$. Let $G_n^*$ be the normalized second moment of the $n$-dimensional sphere $\mathcal{B}(R_l^T)$ of radius $R_l^T$. We have the identity

$$G_n^* |\mathcal{B}(R_l^T)|^{2/n} = \frac{(R_l^T)^2}{(n+2)}$$

Since $|\mathcal{V}^T| = |\mathcal{B}(R_l^T)|$ when $R_l^T$ is the effective radius of $\Lambda^T$, we have

$$R_l^T = \sqrt{\frac{n+2}{n}\frac{G_n^*}{G(\Lambda^T)}}\sqrt{n\sigma^2(\Lambda^T)}.$$

Thus, the covering radius $R_u^T$ of the lattice $\Lambda^T$ is given by

$$R_u^T = \rho_{cov}(\Lambda^T)\sqrt{\frac{n+2}{n}\frac{G_n^*}{G(\Lambda^T)}}\sqrt{n\sigma^2(\Lambda^T)} \tag{16}$$

This expression together with our choice in (15), yields

$$\frac{1}{n}\|\mathbf{x}_s^{(w)}\|^2 \leq P.$$

$\square$

**Relays:** The relay node $i$ receives the signal $\mathbf{y}_i$. As explained earlier, the QMF strategy at the relay is to quantize the received signal using a lattice quantizer and then mapping it to a lattice transmit codebook. The main task is to design the appropriate lattice-to-lattice map that we described informally earlier.

**Quantize:** The signal $\mathbf{y}_i$ is first quantized by using a nested lattice codebook $\Lambda_{Q,i}^*$ which is randomly and independently generated at each relay $i$ by using the nested lattice construction of Section III using the following parameters (same for all relays): Let

$$D_s = \max_i \sum_{j \in \mathcal{M}_{l-1}} |h_{ij}|^2 P. \tag{17}$$

The coarse lattice $\Lambda^Q$ is a scaled version of the lattice $\Lambda$ such that

$$\sigma^2(\Lambda^Q) = 2\eta(D_s + 1) \tag{18}$$

for a constant $\eta > 1$ which is more precisely specified in the proof of Lemma 5.1. Recall that we had set the noise variance to be 1. We denote the generator matrix of the scaled coarse lattice $\Lambda^Q$ by $G_{\Lambda^Q}$. The parameters $m_r$ and $p_r$ are chosen such that $m_r = (\log n)^2$ and $p_r$ is the prime number such that[9]

$$p_r^{m_r} = e^{nR_r}, \quad \text{where} \quad R_r = \frac{1}{2}\log\sigma^2(\Lambda^Q). \tag{19}$$

---

[9] To be more precise, one can take $p_r$ to be the largest prime number such that $p_r \leq e^{nR_r/m_r}$ in which case the rate of the code is $\frac{1}{n}\log p_r^{m_r} \leq R_r$. When $n$ is large, the difference becomes negligible and is therefore ignored.

Note that since $R_r$ is independent of $n$, $p_r = e^{\frac{nR_r}{(\log n)^2}}$, i.e, $p_r \to \infty$ as $n \to \infty$. With the choice in (19) for $R_r$, the second moment of $\Lambda_1^Q$ is given by

$$\sigma^2(\Lambda_1^Q) = \frac{G(\Lambda_1^Q)}{G(\Lambda^Q)}, \tag{20}$$

which follows from (2) by noting that $|\mathcal{V}_1^Q| = |\mathcal{V}^Q|/e^{nR_r}$. It is shown in [19] that the construction of Section III yields nested lattices where the fine lattice is Rogers and Poltyrev-good with high probability if $m \geq (\log n)^2$. (The coarse lattice is both Rogers and Poltyrev-good by construction.) Since both $\Lambda_1^Q$ and $\Lambda^Q$ are Rogers-good w.h.p., $\sigma^2(\Lambda_1^Q) \to 1$ when $n$ increases. Therefore, we are effectively quantizing at the noise level.

At each relay, we independently generate a fine lattice $\Lambda_1^Q$ from the above ensemble, denoted by $\Lambda_{1,i}^Q$, and use the corresponding nested lattice codebook denoted by $\Lambda_{Q,i}^*$. As before, we can index the $e^{nR_r}$ codewords of $\Lambda_{Q,i}^*$ as $\hat{\mathbf{y}}_i^{(k_i)}$, $k_i \in \{1, \ldots, e^{nR_r}\}$ where $k_i$ enumerate the $p_r^{m_r}$ vectors $\mathbf{w}$ in $\mathbb{Z}_{p_r}^{m_r}$ underlying the construction of the nested lattice codebook in (5). Note that by Proposition 3.1, for two indices $k_i \neq k'_i$, $\hat{\mathbf{y}}_i^{(k_i)}$ and $\hat{\mathbf{y}}_i^{(k'_i)}$ are independent, each uniformly distributed over the set of lattice points $p_r^{-1}\Lambda^Q \cap \mathcal{V}^Q$. Moreover, for different relays $i \neq j$, $\hat{\mathbf{y}}_i^{(k_i)}$ and $\hat{\mathbf{y}}_i^{(k_j)}$ are independent.

The quantized signal at relay $i$ is given by

$$\hat{\mathbf{y}}_i = Q_{\Lambda_{1,i}^Q}(\mathbf{y}_i + \mathbf{u}_i) \mod \Lambda^Q$$

where $\mathbf{u}_i$ is a random dither known at the destination node and uniformly distributed over the Voronoi region $\mathcal{V}_{1,i}^Q$ of the fine lattice $\Lambda_{1,i}^Q$. The dithers $\mathbf{u}_i$ are independent for different nodes. We will either say that $\mathbf{y}_i$ is quantized to $\hat{\mathbf{y}}_i$ or to $k_i$ meaning that $\hat{\mathbf{y}}_i = \hat{\mathbf{y}}_i^{(k_i)}$.

**Map and Forward:** Let us scale the coarse lattice $\Lambda$ such that its second moment $\sigma^2(\Lambda^T)$ is given by (15). Let $G_{\Lambda^T}$ denote the generator matrix of the scaled coarse lattice. The quantized signal $\hat{\mathbf{y}}_i$ at relay $i$ is mapped to the transmitted signal $\mathbf{x}_i$ by the following mapping,

$$\mathbf{x}_i = G_{\Lambda^T} \, p_r^{-1} \left( G_i \, p_r \left( G_{\Lambda^Q}^{-1} \, \hat{\mathbf{y}}_i \mod \mathbb{Z}^n \right) \mod p_r \mathbb{Z}^n \right) + \mathbf{v}_i \mod \Lambda^T, \tag{21}$$

where $G_i$ is an $n \times n$ random matrix with its entries uniformly and independently distributed in $0, 1, \ldots, p_r - 1$ and $\mathbf{v}_i$ is a random vector uniformly distributed over $p_r^{-1}\Lambda^T \cap \mathcal{V}^T$, where $\mathcal{V}^T$ is the Voronoi region of $\Lambda^T$. $G_i$ and $\mathbf{v}_i$ are independent for different relay nodes. We denote by $\mathbf{x}_i^{(k_i)}$, $k_i \in \{1, \ldots, e^{nR_r}\}$ the corresponding sequence that the codeword $\hat{\mathbf{y}}_i^{(k_i)}$ is mapped to in (21).

The mapping in (21) can be simplified to the form,

$$\mathbf{x}_i = G_{\Lambda^T} \, G_i \, G_{\Lambda^Q}^{-1} \, \hat{\mathbf{y}}_i + \mathbf{v}_i \mod \Lambda^T. \tag{22}$$

Effectively, it takes the quantization codebook $\Lambda_{Q,i}^*$, expands it by multiplying with a random matrix with large entries (of the order of $p_r$) and then folds it to the Voronoi region of $\Lambda^T$. Since the entries of $G_i$ are potentially very large, even if two codewords are close in $\Lambda_{Q,i}^*$, they are mapped independently to the codewords of the transmit codebook. Note that the complexity of the mapping is polynomial in $n$, while random mapping of the form in [1] has exponential complexity in $n$.

*Proposition 4.2:* The mapping in (21) or (22) has the following properties:

(i) At each relay $i$, the transmitted sequences $\mathbf{x}_i \in \Lambda_i^*$, where $\Lambda_i^*$ is a random nested lattice codebook.

(ii) Given two quantization codewords $\hat{\mathbf{y}}_i^{(k_i)}, \hat{\mathbf{y}}_i^{(k_i')} \in \Lambda_{Q,i}^*$ at relay $i$ such that $k_i \neq k_i'$, the corresponding transmit codewords $\mathbf{x}_i^{(k_i)}$ and $\mathbf{x}_i^{(k_i')}$ are independent, each uniformly distributed over $p_r^{-1}\Lambda^T \cap \mathcal{V}^T$.

(iii) The mapping induces an independent distribution across the relays. Formally, given a set of quantization codewords $\{\hat{\mathbf{y}}_i^{(k_i)}, i \in \mathcal{M}\}$ the corresponding transmit codewords $\{\hat{\mathbf{x}}_i^{(k_i)}, i \in \mathcal{M}\}$ are independently distributed.

*Proof of Proposition 4.2:* The proposition says that the quantization codebooks at each relay are independently mapped to a random nested lattice codebook from the ensemble constructed in the earlier section. The proof is based on the bijection given in (8): There is one-to-one correspondence between the codebook $\Lambda_{Q,i}^*$ and its underlying finite field codebook $\mathcal{C}_{Q,i}$. The mapping in (21) first takes the codeword $\hat{\mathbf{y}}_i \in \Lambda_{Q,i}^*$ to its corresponding codeword in $\mathcal{C}_{Q,i}$. Note that

$$\begin{aligned}
\hat{\mathbf{y}}_i \in \Lambda_{1,i}^Q & \Rightarrow & G_\Lambda^{-1} \hat{\mathbf{y}}_i & \in p_r^{-1}\mathbb{Z}^n \\
& \Rightarrow & G_\Lambda^{-1} \hat{\mathbf{y}}_i \mod \mathbb{Z}^n & \in p_r^{-1}\mathbb{Z}^n \cap [0,1)^n \\
& \Rightarrow & p_r \left( G_\Lambda^{-1} \hat{\mathbf{y}}_i \mod \mathbb{Z}^n \right) & \in \mathbb{Z}_p^n.
\end{aligned}$$

Therefore, $\mathbf{c} = p_r \left( G_\Lambda^{-1} \hat{\mathbf{y}}_i \mod \mathbb{Z}^n \right) \in \mathcal{C}_{Q,i}$. This codeword $\mathbf{c} \in \mathcal{C}_{Q,i}$ is then mapped to a random finite-field codebook $\mathcal{C}_i = \{\mathbf{c}' : \mathbf{c}' = G_i \cdot \mathbf{c}, \mathbf{c} \in \mathcal{C}_{Q,i}\}$. We finally form the nested lattice codebook $\Lambda_i^*$ corresponding to $\mathcal{C}_i$ following again the construction of Section III. Note that, for $\mathbf{c}' \in \mathcal{C}_i$,

$$G_{\Lambda^T} p_r^{-1} \mathbf{c}' + \mathbf{v}_i \mod \Lambda^T \in \Lambda_i^*,$$

where $\Lambda_i^* = (\mathbf{v}_i + \Lambda_{1,i}^T) \mod \Lambda^T$ and $\Lambda_{1,i}^T$ is the fine lattice generated by $\mathcal{C}_i$. Therefore, since $\Lambda_i^*$ is obtained by the construction of Section III from the random linear code $\mathcal{C}_i$, we obtain the result specified in *(i)*. The second property *(ii)* follows by similar observations as in Section III: The random matrix $G_i$ maps every nonzero vector $\mathbf{c} \in \mathcal{C}_{Q,i}$ uniformly at random to another finite field vector in $\mathbb{Z}_p^n$. Two quantized values $\hat{\mathbf{y}}_i^{(k_i)}, \hat{\mathbf{y}}_i^{(k_i')} \in \Lambda_{Q,i}^*$ at relay $i$ such that $k_i \neq k_i'$ correspond to two distinct codewords in $\mathcal{C}_{Q,i}$ which are randomly mapped into new finite field codewords by the random linear map $G_i$. The fact that the lattice points $\mathbf{x}_i^{(k_i)}, \mathbf{x}_i^{(k_i')}$ corresponding to these new finite-field codewords are independently and uniformly distributed over $p_r^{-1}\Lambda^T \cap \mathcal{V}^T$ can be shown by following the arguments in the second part of Proposition 3.1. The third property follows from the independence of the $G_i$'s and $\mathbf{v}_i$'s for different nodes $i$. $\square$

**Destination:** Given its received signal $\mathbf{y}_d$, together with the knowledge of all codebooks, mappings, dithers and channel gains, the decoder performs a consistency check to recover the transmitted message. For each relay $i$ and quantization codeword $\hat{\mathbf{y}}_i^{(k_i)}$, it first forms the signals

$$\tilde{\mathbf{y}}_i^{(k_i)} = \hat{\mathbf{y}}_i^{(k_i)} - \mathbf{u}_i \mod \Lambda^Q. \tag{23}$$

If $\mathbf{y}_i$ denotes the received signal at node $i \in \mathcal{M}_l$ in the $l^{th}$ layer, where $\mathcal{M}_l$ refers to the nodes in the $l^{th}$ layer of the layered network, $\hat{\mathbf{y}}_i$ its quantized version and the $\tilde{\mathbf{y}}_i$ the resultant signal after the transformation above, we

have

$$\tilde{\mathbf{y}}_i = \hat{\mathbf{y}}_i - \mathbf{u}_i \mod \Lambda^Q$$
$$= Q_{\Lambda_{1,i}^Q}(\mathbf{y}_i + \mathbf{u}_i) - \mathbf{u}_i \mod \Lambda^Q$$
$$\stackrel{(a)}{=} (\mathbf{y}_i - \underbrace{(\mathbf{y}_i + \mathbf{u}_i) \mod \Lambda_{1,i}^Q}_{\mathbf{u}'_i}) \mod \Lambda^Q$$
$$= \sum_{j \in \mathcal{M}_{l-1}} h_{ij}\mathbf{x}_j + \mathbf{z}_i - \mathbf{u}'_i \mod \Lambda^Q, \qquad (24)$$

where (a) follows by definition in (7) and the quantization error $\mathbf{u}'_i = (\mathbf{y}_i + \mathbf{u}_i) \mod \Lambda_{1,i}^Q$ is independent of $\mathbf{y}_i$ and is uniform over the Voronoi region of $\Lambda_{1,i}^Q$. This follows by the so called Crypto Lemma which is extensively used in the sequel. We state the lemma below for completeness.

*Lemma 4.1 (Crypto Lemma,[15]):* Let $\mathbf{u}$ be a random variable uniformly distributed over the Voronoi region $\mathcal{V}$ of a lattice $\Lambda$. For any random variable $\mathbf{x} \in \mathcal{V}$, statistically independent of $u$, we have the sum $\mathbf{y} = \mathbf{x} + \mathbf{u} \mod \Lambda$ is uniformly distributed over $\mathcal{V}$, and is statistically independent of $\mathbf{x}$.

To conclude that $\mathbf{u}'_i = (\mathbf{y}_i + \mathbf{u}_i) \mod \Lambda_{1,i}^Q$ is independent of $\mathbf{y}_i$, note that $\mathbf{u}'_i = (\mathbf{y}_i \mod \Lambda_{1,i}^Q + \mathbf{u}_i) \mod \Lambda_{1,i}^Q$. By the Crypto Lemma, $\mathbf{u}'_i$ is independent of $\mathbf{y}_i \mod \Lambda_{1,i}^Q$. Since it also independent of $Q_{\Lambda_{1,i}^Q}(\mathbf{y}_i)$, we conclude that $\mathbf{u}'_i$ is independent of $\mathbf{y}_i$.

The decoder then forms the set $\hat{\mathcal{W}}$ of messages $\hat{w}$ such that

$$\hat{\mathcal{W}} = \{\hat{w} : \exists \{k_i\} \text{such that } (\mathbf{x}_s^{(\hat{w})}, \mathbf{y}_d, \{\tilde{\mathbf{y}}_i^{(k_i)}, \mathbf{x}_i^{(k_i)}\}_{i \in \mathcal{M}}) \in \widetilde{\mathcal{A}}_\epsilon\} \qquad (25)$$

where $\widetilde{\mathcal{A}}_\epsilon$ denotes consistency. We define consistency as follows: For a given set of indices $\{k_i\}_{i \in \mathcal{M}}$, we say $(\mathbf{x}_s^{(\hat{w})}, \mathbf{y}_d, \{\tilde{\mathbf{y}}_i^{(k_i)}, \mathbf{x}_i^{(k_i)}\}_{i \in \mathcal{M}}) \in \widetilde{\mathcal{A}}_\epsilon$ if

$$\|(\tilde{\mathbf{y}}_i^{(k_i)} - \sum_{j \in \mathcal{M}_{l-1}} h_{ij}\mathbf{x}_j^{(k_j)}) \mod \Lambda^Q\|^2 \leq n\,\sigma_c^2, \qquad (26)$$

for all $i \in \mathcal{M}_l$, $1 \leq l \leq l_d$ where for convenience of notation we have denoted $\mathbf{x}_s^{(\hat{w})} = \mathbf{x}_j^{(k_j)}, j \in \mathcal{M}_0$, and $\mathbf{y}_d = \tilde{\mathbf{y}}_i^{(k_i)}, i \in \mathcal{M}_{l_d}$. Recall that $\mathcal{M}_l$ refers to the nodes in the $l^{th}$ layer of the layered network. We choose

$$\sigma_c^2 = 2(1 + \epsilon) \qquad (27)$$

for a constant $\epsilon > 0$ that can be taken arbitrarily small. Recall from (1), (18) that the noise variance and the quantization error were set to 1.

The decoder declares $\hat{w}$ to be the transmitted message if it is the unique message in $\hat{\mathcal{W}}$. An error occurs when the declared message $\hat{w}$ is not the same as $w$, or when there are multiple messages in $\hat{\mathcal{W}}$.

We can interpret the consistency check as follows: For each layer $l = 1, \ldots, l_d - 1$ the decoder picks a set of potential (quantized) received sequences $\{\tilde{\mathbf{y}}_i^{(k_i)}\}_{i \in \mathcal{M}_l}$ and the transmit sequences corresponding to them $\{\mathbf{x}_i^{(k_i)}\}_{i \in \mathcal{M}_l}$. It checks for each layer $l$, whether the inputs and outputs are consistent, or jointly "typical", i.e., whether the examined outputs $\{\tilde{\mathbf{y}}_i^{(k_i)}\}_{i \in \mathcal{M}_l}$ at the layer $l$ can be explained (to within the noise and quantization error) by the transmitted sequences $\{\mathbf{x}_i^{(k_i)}\}_{i\in\mathcal{M}_{l-1}}$ of layer $l-1$ for indices $\{k_i\}$. The relation (24) and the fact that $\Lambda_{1,i}^Q$ is Rogers-good ensures that for large $n$ the inputs $\{\mathbf{x}_i^{(k_i)}\}_{i\in\mathcal{M}_{l-1}}$ and those outputs $\{\tilde{\mathbf{y}}_i^{(k_i)}\}_{i\in\mathcal{M}_l}$ that are generated from these inputs are consistent with high probability. Note that the termination conditions for the consistency check across the layers are known, i.e., $\mathbf{x}_s$ is known for the message being tested, and $\mathbf{y}_d$ is the observed sequence at the destination. Therefore, effectively the decoder checks whether there exists a plausible set of input and output sequences at each relay that under the message $w$ could yield the observation $\mathbf{y}_d$. Note that the definition of consistency in (26) is closely related to weak typicality. Indeed, it is a variant of the weak typicality condition for Gaussian vectors. Therefore, effectively our decoder is a typicality decoder designed for lattices.

## A. Multiple Antennas

A slightly modified version of the above scheme applies to the case of multiple transmit and receive antennas at each node. Let $M_i$ be the number of transmit and $N_i$ be the number of receive antennas at each node.

**Source:** The source node $s$ maps its message to $M_s$ independent nested lattice codebooks $\Lambda_1^*, \ldots \Lambda_{M_s}^*$ and transmits its codeword from its corresponding transmit antenna.

**Relays:** The relay node $i$ receives $N_i$ signals denoted $\mathbf{y}_{i,1}, \ldots, \mathbf{y}_{i,N_i}$. It individually quantizes each signal by adding an independent random dither,

$$\hat{\mathbf{y}}_{i,a} = Q_{\Lambda_{1,i}^Q}(\mathbf{y}_{i,a} + \mathbf{u}_{i,a}) \mod \Lambda^Q, \qquad a = 1, \ldots, N_i.$$

The transmitted codeword from the $b$'th transmit antenna of node $i$ is given by

$$\mathbf{x}_{i,b} = G_{\Lambda^T} \sum_{l=1}^{N_i} G_{i,b,a} G_{\Lambda^Q}^{-1} \hat{\mathbf{y}}_{i,a} + \mathbf{v}_{i,b} \mod \Lambda^T. \qquad (28)$$

where $G_{i,b,a}$ is $n\times n$ random matrix independent across $i$, $a$ and $b$. The mapping is modified from (22) so that at each relay, the set of quantization codewords $\hat{\mathbf{y}}_{i,1}^{(k_{i,1})}, \ldots, \hat{\mathbf{y}}_{i,N_i}^{(k_{i,N_i})}$ is mapped independently to $M_i$ random nested lattice codebooks. For each of the $M_i$ random codebooks, two different sets of quantization codewords $\hat{\mathbf{y}}_{i,1}^{(k_{i,1})}, \ldots, \hat{\mathbf{y}}_{i,N_i}^{(k_{i,N_i})}$ and $\hat{\mathbf{y}}_{i,1}^{(k'_{i,1})}, \ldots, \hat{\mathbf{y}}_{i,N_i}^{(k'_{i,N_i})}$ are mapped uniformly and independently to the set $p_r^{-1}\Lambda^T \cap \mathcal{V}^T$, if $\exists a \in 1, \ldots, N_i$ such that $k_{i,a} \neq k'_{i,a}$.

**Destination:** Similarly to the single antenna case, for a given message $\hat{w}$ and a set of observations $\mathbf{y}_{d,1}, \ldots, \mathbf{y}_{d,N_d}$, the destination node checks whether there exist a set of indices $\{k_{i,a}\}_{i\in\mathcal{M}, 1\le a \le N_i}$ such that the inputs and outputs at each layer are consistent.

The error analysis in the next section is performed for the single antenna case and follows similar lines for the case of multiple antennas.

## V. ERROR ANALYSIS

Due to the nature of the decoder at the destination, described in (25), an error occurs when either the transmitted message $w$ is not in $\hat{\mathcal{W}}$ or when there is a message $w' \neq w$ which is in $\hat{\mathcal{W}}$. The transmitted message $w$ from the source and the resulting observation at the destination will pass the consistency check in (26) with high probability because the channel and the quantization noise at relays will be typical, confined inside a ball of radius $\sqrt{n}\sigma_c$, with probability approaching 1 as $n$ increases. This is made more precise in (33). An error occurs when there exists an incorrect message $w'$ that is also consistent with the observation at the destination, *i.e.,* there exists a plausible sequence of received (quantized) values that can result in the signal seen at the destination if $w'$ were transmitted.

Te main focus in the error analysis is on bounding the probability that a particular incorrect message $w'$ will pass the check when $w$ is transmitted. We first split this error event into $2^N$ disjoint subevents indexed by $\Omega$. Consider the two plausible sequences of received (quantized) values that correspond to $w$ and $w'$. $\Omega$ denotes the event that these two sequences are different for nodes in the set $\Omega$ and same for nodes in $\Omega^c$. When this is the case, we say that nodes in $\Omega$ can "distinguish" between the correct and the incorrect message while the nodes in $\Omega^c$ can not. This notion of distinguishability was also used in [1][10]. The probability of $\Omega$ can be split into parts: the probability that the nodes in $\Omega^c$ are confused times the probability that the nodes in $\Omega$ are not confused given that the nodes in $\Omega^c$ are confused. We upper bound the first probability in Lemma 5.2 and the second probability in Lemma 5.3. Combining the results of the two lemmas we obtain the conclusion in Theorem 2.2.

Let $w$ be the transmitted message from the source. As described earlier, We will analyze error event:

$$\mathcal{E} \stackrel{def}{=} \left\{ w \notin \hat{\mathcal{W}} \right\} \cup \left\{ w' \in \hat{\mathcal{W}} \text{ for some } w' \neq w \right\}, \tag{29}$$

where $\hat{\mathcal{W}}$ is defined in (25). If $w$ is the transmitted message, this probability can be upper bounded as,

$$\mathbb{P}[\mathcal{E}] \leq e^{nR} \, \mathbb{P}[w' \in \hat{\mathcal{W}}, w' \neq w] + \underbrace{\mathbb{P}[w \notin \hat{\mathcal{W}}]}_{<\epsilon} \tag{30}$$

where $\mathbb{P}[w' \in \hat{\mathcal{W}}]$ is the probability that a particular incorrect message $w' \neq w$ passes the consistency check in (26). This probability can be upper bounded by using the union bound as

$$\mathbb{P}\left[\exists \{k'_i\}_{i \in \mathcal{M}} \text{ s.t. } (\mathbf{x}_s^{(w')}, \mathbf{y}_d, \{\tilde{\mathbf{y}}_i^{(k'_i)}, \mathbf{x}_i^{(k'_i)}\}_{i \in \mathcal{M}}) \in \widetilde{\mathcal{A}}_\epsilon \right] \leq \sum_{k'_1,\ldots,k'_N} \mathbb{P}\left[(\mathbf{x}_s^{(w')}, \mathbf{y}_d, \{\tilde{\mathbf{y}}_i^{(k'_i)}, \mathbf{x}_i^{(k'_i)}\}_{i \in \mathcal{M}}) \in \widetilde{\mathcal{A}}_\epsilon \right], \tag{31}$$

where each term in the summation is the probability that the corresponding set of particular quantization indices $k'_1, \ldots, k'_N$ make $w'$ plausible with the observation at the destination.

The second term $\mathbb{P}[w \notin \hat{\mathcal{W}}]$ in (30) is small for large $n$ since for the correct message the consistency check in (26) simply reduces to checking whether the quantization and the additive noise are typical. Let $\{k_1, \ldots, k_N\}$ be the quantization indices produced during transmission of $w$. The consistency check in (26) for these actual quantization codewords is given by

$$\|(\mathbf{z}_i - \mathbf{u}'_i) \mod \Lambda^Q\|^2 \leq n\,\sigma_c^2, \tag{32}$$

for all $i \in \mathcal{M}$ where we used the relation (24). The noise $\mathbf{z}_i$ is $\mathcal{N}(0,1)$, therefore for large $n$, $\mathbb{P}[\|\mathbf{z}_i\|^2 \leq n(1+\epsilon)] \to 1$. This can observed from Lemma 5.2. On the other hand, the quantization noise $\mathbf{u}'_i$ is uniformly distributed over the Voronoi region of $\Lambda_{1,i}^Q$. Since this lattice is Roger's good, its covering radius $R_u \to \sigma^2(\Lambda_{1,i}^Q) \to 1$ when $n$ is large. Therefore $\mathbb{P}[\|\mathbf{u}_i'\|^2 \leq n(1+\epsilon)] \to 1$. This can be verified by combining the results of Lemma 7.1 and Lemma 7.2. Since $\|\mathbf{z}_i - \mathbf{u}_i'\|^2 \leq \|\mathbf{z}_i\|^2 + \|\mathbf{u}_i'\|^2$, we conclude that $\mathbb{P}[\|(\mathbf{z}_i - \mathbf{u}_i')\|^2 \leq 2(1+\epsilon)n] \to 1$. Since there are finitely many of relays, the union bound gives the same conclusion simultaneously for all relays. Therefore, we conclude that

$$\mathbb{P}[w \notin \hat{\mathcal{W}}] \to 1, \tag{33}$$

for large $n$. In the above argument, we have ignored the $\mod \Lambda^Q$ operation in (32) because $\mathbf{z}_i - \mathbf{u}_i'$ lies in the Voronoi region of $\Lambda^Q$ with high probability due to our choice for $\sigma^2(\Lambda^Q)$ and the fact that the lattices are Roger's good.

---

[10]In [1] this was done for Gaussian transmit codebooks and scalar quantizers, whereas in this paper we used lattice vector quantizers and lattice transmit codebooks.

In order to compute the upper bound in (31), we will condition on the event that the correct message $w$ produced a sequence of indices $k_1, \ldots, k_N$. Since these are generic indices, we can carry out the entire calculation conditioned on a particular sequence $k_1, \ldots, k_N$ and then average over it. In this case, the summation over the $N$ indices $k_1', \ldots, k_N'$ in (31) can be rearranged to yield

$$\sum_{\Omega} \sum_{\substack{k_i', i \in \Omega \\ k_i' \neq k_i}} \underbrace{\mathbb{P}\Big((\mathbf{x}_s^{(w')}, \mathbf{y}_d, \{\tilde{\mathbf{y}}_i^{(k_i')}, \mathbf{x}_i^{(k_i')}\}_{i \in \mathcal{M}}) \in \widetilde{\mathcal{A}}_\epsilon \text{ s.t. } k_i' = k_i, \ i \in \Omega^c\Big)}_{\mathcal{P}}, \tag{34}$$

where $\Omega \subset \mathcal{M}$ is a source-destination cut of the network, *i.e.*,

$$\Omega \subset \mathcal{M} \text{ such that } s \in \Omega, d \in \Omega^c. \tag{35}$$

Now, let us examine the probability denoted by $\mathcal{P}$. For a given set of $\{k_i'\}_{i \in \mathcal{M}}$ such that $k_i' = k_i, \ i \in \Omega^c$ and $k_i' \neq k_i, \ i \in \Omega$, the consistency condition for a node $i \in \mathcal{M}_l$ in the $l$th layer of the network is given by (26) as

$$\|(\tilde{\mathbf{y}}_i^{(k_i')} - \sum_{j \in \mathcal{M}_{l-1}} h_{ij} \mathbf{x}_j^{(k_j')}) \mod \Lambda^Q\|^2 \leq n \sigma_c^2, \qquad \forall i \in \mathcal{M}_l, 1 \leq l \leq l_d \tag{36}$$

where for convenience of notation we denote $\mathbf{y}_d = \tilde{\mathbf{y}}_i^{(k_i)}, i \in \mathcal{M}_{l_d}$ and $\mathbf{x}_s^{(w')} = \mathbf{x}_j^{(k_j')}, j \in \mathcal{M}_0$. The condition in (36) takes two different forms depending on whether $i \in \Omega$ or $i \in \Omega^c$:

For nodes $i \in \Omega^c$, $\tilde{\mathbf{y}}_i^{(k_i')} = \tilde{\mathbf{y}}_i^{(k_i)}$ and from (24) it is related to the inputs from the previous layer as

$$\tilde{\mathbf{y}}_i^{(k_i)} = \sum_{j \in \mathcal{M}_{l-1}} h_{ij} \mathbf{x}_j^{(k_j)} + \mathbf{z}_i - \mathbf{u}_i' \mod \Lambda^Q. \tag{37}$$

In this case, the condition (36) is equivalent to

$$\mathcal{A}_i = \{\|(\sum_{j \in \Omega_{l-1}} h_{ij}(\mathbf{x}_j^{(k_j)} - \mathbf{x}_j^{(k_j')}) + \mathbf{z}_i - \mathbf{u}_i') \mod \Lambda^Q\|^2 \leq n \sigma_c^2\}, \tag{38}$$

where $\Omega_{l-1} = \Omega \cap \mathcal{M}_{l-1}$ and we denote this event by $\mathcal{A}_i$[11]. Note that we have have used the fact that for nodes $i \in \Omega^c$, since $k_i' = k_i$, we have $\mathbf{x}_i^{(k_i')} = \mathbf{x}_i^{(k_i)}$.

---

[11]The condition is slightly different for the destination node $d$, in particular it does not contain the term $\mathbf{u}_i'$ in (38), since we operate directly on the observation $y_d$ and not it's quantized version. This fact is ignored since it does not create any significant difference in the below analysis. Alternatively, it can be assumed that the destination node first quantizes its received signal and then performs the consistency check.

DRAFT

For nodes $i \in \Omega$, the condition yields

$$\mathcal{B}_i = \{\|(\tilde{\mathbf{y}}_i^{(k'_i)} - \sum_{j \in \Omega^c_{l-1}} h_{ij}\mathbf{x}_j^{(k_j)} - \sum_{j \in \Omega_{l-1}} h_{ij}\mathbf{x}_j^{(k'_j)}) \mod \Lambda^Q\|^2 \leq n\,\sigma_c^2\}, \tag{39}$$

where $\Omega^c_{l-1} = \Omega^c \cap \mathcal{M}_{l-1}$ and we denote this event by $\mathcal{B}_i$).

To summarize, for $i \in \Omega^c$, $\mathcal{A}_i$ is the event that $\tilde{\mathbf{y}}_i^{(k'_i)}$ is consistent (jointly typical) with transmitted sequences corresponding to $\{k'_i\}$, and $\mathcal{B}_i$ is the corresponding event for nodes $i \in \Omega$.

Now, coming back to the calculation of $\mathcal{P}$ in (34), we can write

$$\begin{aligned}\mathcal{P} &= \mathbb{P}\left(\{\mathcal{A}_i, i \in \Omega^c\}, \{\mathcal{B}_i, i \in \Omega\}\right) \\ &= \mathbb{P}\left(\mathcal{A}_i, i \in \Omega^c\right) \mathbb{P}\left(\mathcal{B}_i, i \in \Omega \,|\, \mathcal{A}_i, i \in \Omega^c\right).\end{aligned} \tag{40}$$

Note that due to Proposition 4.2, for all $j \in \mathcal{M}$, when $k'_j \neq k_j$, the relay mapping induces transmit sequences $\mathbf{x}_j^{(k_j)}, \mathbf{x}_j^{(k'_j)}$ that are pairwise independent and uniformly distributed over $p_r^{-1}\Lambda^T \cap \mathcal{V}^T$.[12] Also, due to the dithering in (23), $\tilde{\mathbf{y}}_i^{(k'_i)}$ in (39) is uniformly distributed over the Voronoi region $\mathcal{V}_1^Q$ of the quantization lattice point $\hat{\mathbf{y}}_i^{(k'_i)}$.

We will first bound the probability $\mathbb{P}\left(\mathcal{A}_i, i \in \Omega^c\right)$ by conditioning on the event defined in the following lemma, which is proved in the Appendix.

*Lemma 5.1:* Let us define the following event,

$$\mathcal{E}_1 \stackrel{def}{=} \left\{\exists\, i \in \mathcal{M}, \exists\, \{k_j, k'_j\} \text{ s.t. } \sum_j h_{ij}(\mathbf{x}_j^{(k_j)} - \mathbf{x}_j^{(k'_j)}) + \mathbf{z}_i - \mathbf{u}'_i \notin \mathcal{V}^Q\right\}, \tag{41}$$

then we have $\mathbb{P}(\mathcal{E}_1) \to 0$ as $n \to \infty$.

When $\mathcal{E}_1$ is true, we declare this as an error. This adds a vanishing term to the decoding error probability by the above lemma. Conditioning on the complement of $\mathcal{E}_1$ allows us to get rid of the mod operation w.r.t $\Lambda^Q$ in (38). Given $\mathcal{E}_1^c$, the event $\mathcal{A}_i$, for $i \in \Omega^c$ is equivalent to

$$\mathcal{A}'_i = \{\|(\sum_{j \in \Omega_{l-1}} h_{ij}(\mathbf{x}_j^{(k_j)} - \mathbf{x}_j^{(k'_j)})) + \mathbf{z}_i - \mathbf{u}'_i\|^2 \leq n\,\sigma_c^2\}. \tag{42}$$

Therefore, we have

$$\begin{aligned}\mathbb{P}\left(\mathcal{A}_i, i \in \Omega^c\right) &= \mathbb{P}\left(\mathcal{E}_1^c\right)\mathbb{P}\left(\mathcal{A}_i(\{k'_i\}), i \in \Omega^c \,|\, \mathcal{E}_1^c\right) + \mathbb{P}\left(\mathcal{E}_1\right)\mathbb{P}\left(\mathcal{A}_i(\{k'_i\}), i \in \Omega^c \,|\, \mathcal{E}_1\right) \\ &\leq \mathbb{P}\left(\mathcal{A}_i, i \in \Omega^c, \mathcal{E}_1^c\right) + \mathbb{P}\left(\mathcal{E}_1\right) = \mathbb{P}\left(\mathcal{A}'_i, i \in \Omega^c, \mathcal{E}_1^c\right) + \mathbb{P}\left(\mathcal{E}_1\right) \\ &\leq \mathbb{P}\left(\mathcal{A}'_i, i \in \Omega^c\right) + \underbrace{\mathbb{P}\left(\mathcal{E}_1\right)}_{\to 0} \stackrel{n \to \infty}{=} \mathbb{P}\left(\mathcal{A}'_i, i \in \Omega^c\right)\end{aligned} \tag{43}$$

We upperbound this probability in the following lemma.

---

[12]For the source node, $\mathbf{x}_j^{(k_j)}$ and $\mathbf{x}_j^{(k'_j)}$ or equivalently $\mathbf{x}_s^{(w)}$ and $\mathbf{x}_s^{(w')}$ are uniformly distributed over $p^{-1}\Lambda^T \cap \mathcal{V}^T$ where $p$ is different than $p_r$. However, this fact does not create any difference in the following analysis and is therefore ignored.

*Lemma 5.2:*

$$\mathbb{P}\left(\mathcal{A}'_i, i \in \Omega^c\right) = \mathbb{P}\left(\|\sum_{j \in \Omega_{l-1}} h_{ij}(\mathbf{x}_j^{(k_j)} - \mathbf{x}_j^{(k'_j)}) + \mathbf{z}_i - \mathbf{u}'_i\|^2 \leq n\,\sigma_c^2, \forall i \in \Omega^c\right)$$

$$\leq e^{-n\left(I(X_\Omega; HX_\Omega + Z_{\Omega^c}) - \frac{1}{2}|\Omega^c|(1+\log(1+\epsilon)) - o_n(1)\right)},$$

where $X_i, i \in \Omega$ are i.i.d Gaussian random variables $\mathcal{N}(0, P)$, $Z_{\Omega^c}$ are i.i.d Gaussian random variables $\mathcal{N}(0, \sigma^2)$ and $H$ is the channel transfer matrix from nodes in $\Omega$ to nodes in $\Omega^c$.

The proof of the lemma involves two main steps. Recall that $\mathbf{x}_j^{(k_j)}, \mathbf{x}_j^{(k'_j)}, j \in \Omega$ are elements of a lattice and therefore are discrete random variables, which are uniformly distributed over $p_r^{-1}\Lambda^T \cap \mathcal{V}^T$ and are pairwise independent. We first show that the probability in the lemma is upper bounded by

$$e^{n\epsilon_2}\mathbb{P}\left[\|\sum_{j \in \Omega_{l-1}} h_{ij}(\mathbf{x}_j - \mathbf{x}'_j) + \mathbf{z}_i - \mathbf{z}'_i\|^2 \leq n\,\sigma_c^2, \forall i \in \Omega^c\right] \tag{44}$$

where $\mathbf{x}_j, \mathbf{x}'_j, j \in \Omega$ and $\mathbf{z}'_i, i \in \Omega^c$ are all independent Gaussian random variables such that $\mathbf{x}_j, \mathbf{x}'_j \sim \mathcal{N}(0, \sigma_x^2 I_n)$, $\mathbf{z}'_i \sim \mathcal{N}(0, \sigma_z^2 I_n)$ and $\sigma_x^2 \to \sigma^2(\Lambda^T) \to P$ as $n \to \infty$ if $\Lambda^T$ is Rogers-good, $\sigma_z^2 \to \sigma^2(\Lambda_{1,i}^Q) \to 1$ as $n \to \infty$ if $\Lambda_{1,i}^Q$ is Rogers-good, which is our case here. $\epsilon_2 \to 0$ when $n$ increases, again if $\Lambda^T$ and $\Lambda_{1,i}^Q$ are Rogers-good. Given this translation to Gaussian distributions the problem becomes very similar to the one for Gaussian codebooks in [1]. The second step is to bound the probability in (44) by following a similar approach to [1]. The proof is given in the Appendix.

Using Lemma 5.2, Lemma 5.1 in (40), we can upperbound the error probability given in (34) as,

$$\sum_\Omega e^{-n\left(I(X_\Omega; HX_\Omega + Z_{\Omega^c}) - \frac{1}{2}|\Omega^c|(1+\log(1+\epsilon)) - o_n(1)\right)} \sum_{\substack{k'_i, i \in \Omega \\ k'_i \neq k_i}} \mathbb{P}\left(\mathcal{B}_i, i \in \Omega \mid \mathcal{A}_i, i \in \Omega^c\right) \tag{45}$$

The last term in (45) is upper bounded in the following lemma.

*Lemma 5.3:* We have

$$\sum_{\substack{k'_i, i \in \Omega \\ k'_i \neq k_i}} \mathbb{P}\left(\mathcal{B}_i, i \in \Omega \mid \mathcal{A}_i, i \in \Omega^c\right) \leq e^{|\Omega|n\frac{1}{2}(\log(2(1+\epsilon)) + 1 + o_n(1))}. \tag{46}$$

The proof of the lemma is based on two steps. We first argue that due to the random construction of the quantization codebook at each relay, $\tilde{\mathbf{y}}_i^{(k'_i)}$ is uniformly distributed over the Voronoi region $\mathcal{V}^Q$ of the quantizer and is independent across different relay nodes $i \in \Omega$. Due to the Crypto Lemma ( Lemma 4.1), this is also true for the random variables

$$\nu_i = \tilde{\mathbf{y}}_i^{(k'_i)} - \sum_{j \in \Omega_{l-1}^c} h_{ij}\mathbf{x}_j^{(k_j)} - \sum_{j \in \Omega_{l-1}} h_{ij}\mathbf{x}_j^{(k'_j)} \mod \Lambda^Q, \quad i \in \Omega$$

appearing in the definition of the event $\mathcal{B}_i$ because the $\tilde{\mathbf{y}}_i$'s and $\mathbf{x}_i$'s are independent of each other. More precisely, due to the random mapping between the quantization and transmission codebooks at each relay, the set of random variables $\{\tilde{\mathbf{y}}_i^{(k'_i)}, i \in \Omega\}$ are independent from the set of random variables $\{\mathbf{x}_i^{(k_i)}, i \in \Omega^c\}, \{\mathbf{x}_i^{(k'_i)}, i \in \Omega\}$. Therefore by the Crypto Lemma [15], $\nu_i$'s are also independent of the $\mathbf{x}_i$'s which allows to remove the conditioning on the event $\mathcal{A}_i, i \in \Omega^c$ in (46), which only governs $x_i$'s. Finally, each term in the summation in (46) reduces to evaluating the probability $\mathbb{P}\left(\|\nu\|^2 \leq n\,\sigma_c^2\right)$, where $\nu$ is a random variable uniformly distributed over $\mathcal{V}^Q$. This probability is upper bounded in the following lemma which is proved in the Appendix.

*Lemma 5.4:* Let $\nu$ be uniformly distributed over $\mathcal{V}^Q$. We have,

$$\mathbb{P}\left(\|\nu\|^2 \leq n\,\sigma_c^2\right) \leq e^{-\frac{n}{2}\left(\log\left(\frac{\sigma^2(\Lambda^Q)}{\sigma_c^2}\right) - 1 + \frac{\sigma_c^2}{\sigma^2(\Lambda^Q)} - o_n(1)\right)}.$$

Using the results of Lemma 5.2 and Lemma 5.3 in (45), together with the summation over all possible source-destination cuts in (34), we obtain

$$\mathbb{P}[w' \in \hat{\mathcal{W}}, w' \neq w] \leq \sum_{\Omega} e^{-n(I(X_\Omega; HX_\Omega + Z_{\Omega^c}) - N - o_n(1))} \leq 2^N e^{-n \min_{\Omega}(I(X_\Omega; HX_\Omega + Z_{\Omega^c}) - (1+(\log 2)/2)N - o_n(1))}. \tag{47}$$

Combining this upper bound with (30), demonstrates that if $R < \min_\Omega I(X_\Omega; HX_\Omega + Z_{\Omega^c}) - (1 + (\log 2)/2)N$ then $\mathbb{P}[\mathcal{E}] \to 0$. This proves the main result of this paper which is stated in Theorem 2.1.[13]

## VI. HALF-DUPLEX RELAY NETWORKS

A common practical constraint in wireless networks is that nodes can not transmit and receive at the same time on the same frequency band, termed as the half-duplex constraint. In this section, we will extend the constant gap result of the earlier sections to half-duplex relay networks.

Since each node in a half-duplex network can be in either transmitting or receiving mode, there are $2^N$ different possible states for the overall network. Each state is a partitioning of the nodes into two distinct sets of transmitters and receivers. A schedule defines the fraction of time the network operates in each of these $2^N$ states. We call a schedule fixed if it is decided ahead of time and revealed to all the nodes in the network. As shown in [1], the quantize-map-and-forward relaying scheme can be combined with a fixed schedule and applied in half-duplex networks. Theorem 8.3 of [1] shows that the rate achieved by the quantize-map-and-forward scheme is within a constant gap to the capacity of the half-duplex network evaluated under fixed schedules and uniform power allocation across different states. However, since the half-duplex schedule can also be random and not fixed, it is not clear if the performance of the quantize-map-and-forward scheme is within a constant gap to the actual information-theoretic capacity of the network. For example, [12] demonstrates that random schedules can yield higher rates than fixed schedules in wireless networks. Even more importantly, the average transmit power constraint allows to optimize the transmit power of each node across the $2^N$ states of the network and not necessarily transmit with the same power at every state. In this section, we improve the result of [1] by showing that the quantize-map-and-forward scheme combined with a fixed schedule and uniform power allocation $P$ across all the states of the network achieves the information-theoretic capacity of the network within $3N$ bits/s/Hz in the single antenna case (or $2\sum_{i \in s, \mathcal{M}} M_i + N$ bits/s/Hz in the case of multiple antennas.) For simplicity, we concentrate on the single-antenna case in the sequel.

---

[13]The gap in Theorem 2.1 is for the complex case.

The multiple-antenna case follows similarly. Our results are based on the memoryless model developed in [12] for half-duplex relay networks.

## A. Half-Duplex Channel Model

We follow the model developed in [12]. Due to the half-duplex constraint each node $i$ in the network can be in either transmit or receive mode, denoted by $m_i = T$ and $m_i = L$ respectively. When $m_i = T$, the received signal of the node $i$ is equal to zero, i.e. $y_i = 0$. When $m_i = L$, the transmitted signal by the node $i$ is equal to zero, i.e. $x_i = 0$. These constraints can be incorporated to the channel model by considering the transmitted signals which are inputs to the channel to be the vectors $\bar{x}_i = (x_i, m_i)$ with alphabet

$$\mathcal{X}_i = \{(0, L), (\mathbb{C}, T)\}$$

where $\mathbb{C}$ is the set of complex numbers. Accordingly, the Gaussian channel model is modified to

$$y_i = \begin{cases} \sum_{j \neq i} H_{ij} x_j + z_i & \text{if } m_i = L \\ 0 & \text{if } m_i = T, \end{cases}$$

where as before $H_{ij}$'s are the corresponding channel matrices and $z_i$ is the additive Gaussian noise. As before, an individual average power constraint applies to each transmitting node $i$, i.e.,

$$\mathbb{E}[||x_i||^2] \leq P, \ \forall i \in \mathcal{M} \cup \{s\},$$

where as we recall from Section II that $\mathcal{M}$ is the set of all relay nodes, excluding the source and the destination nodes. We assume that the source node is always transmitting and the destination nodes are always receiving.

## B. Cut-set Upper Bound

As noted in [12], the memoryless model allows to use the existing theory on memoryless relay networks. In particular, applying the cut-set bound [8, Theorem 14.10.1], we can upper bound the communication rate between the source and the destination in the half duplex network by

$$\overline{C}_{h.d} = \max_{\substack{p_{\bar{x}_{s,\mathcal{M}}}(\cdot) \\ s.t. \, \mathbb{E}[||x_i||^2] \leq P, \forall i}} \min_{\Omega} I(\bar{x}_\Omega; y_{\Omega^c} | \bar{x}_{\Omega^c}) = \max_{\substack{p_{m_{\mathcal{M}}, x_{s,\mathcal{M}}}(\cdot) \\ s.t. \, \mathbb{E}[||x_i||^2] \leq P, \forall i}} \min_{\Omega} I(m_\Omega, x_\Omega; y_{\Omega^c} | m_{\Omega^c}, x_{\Omega^c}), \quad (48)$$

where $\bar{x}_{s,\mathcal{M}} = \{(x_i, m_i), i \in \{s, \mathcal{M}\}\}$, $m_\mathcal{M} = \{m_i, i \in \mathcal{M}\}$, $\Omega$ is a source-destination cut of the network and $\bar{x}_\Omega = \{\bar{x}_i, i \in \Omega\}$, $y_{\Omega^c} = \{y_i, i \in \Omega^c\}$ and $\bar{x}_{\Omega^c}$, $m_\Omega$, $x_\Omega$, $m_{\Omega^c}$, $x_{\Omega^c}$ are defined similarly.

## C. A Simple Upper Bound on the Cut-set Upper Bound

In this section, we develop an upper bound on the cut-set upper bound in (48) that provides the connection to the performance of quantize-map-and-forward with fixed schedules and uniform power allocation. First note that the mutual information in (48) can be separated into two terms,

$$I(m_\Omega, x_\Omega; y_{\Omega^c} | m_{\Omega^c}, x_{\Omega^c}) = I(x_\Omega; y_{\Omega^c} | m_\Omega, m_{\Omega^c}, x_{\Omega^c}), + I(m_\Omega; y_{\Omega^c} | m_{\Omega^c}, x_{\Omega^c}) \quad (49)$$

$$\leq I(x_\Omega; y_{\Omega^c} | m_\mathcal{M}, x_{\Omega^c}) + N. \quad (50)$$

The inequality (50) follows by upper bounding the second mutual information in (49) by $N$ bits/s/Hz since each of the $N$ random variables $m_i, i \in \mathcal{M}$ are binary. The first mutual information governs a fixed schedule and the expression in (48) involves a maximization of this mutual information over all possible schedules. Moreover, we can allocate different transmit powers for the nodes in different states of this optimal schedule. Below we will show that an optimal power allocation across the states differs by at most $2N$ bits/s/Hz from the case where all the nodes transmit with uniform power $P$ whenever they are transmitting. A priori, one can expect this gap to scale with $2^N$, the number of different states of the network.

Let us denote the average transmit power of node $i$ at state $m$ with $P_i(m)$. Clearly the individual power constraint translates to $\mathbb{E}[P_i(m)] \leq P$, where the expectation is over the states. Then, the cutset upper bound can be rewritten and upper bounded as follows:

$$\overline{C}_{h,d} \leq \max_{\substack{p_{m_\mathcal{M}}(\cdot)\, p_{x_{s,\mathcal{M}}|m_\mathcal{M}}(\cdot) \\ s.t.\, \mathbb{E}[||x_i||^2] \leq P, \forall i}} \min_{\Omega} I(x_\Omega; y_{\Omega^c}|m_\mathcal{M}, x_{\Omega^c}) + N$$

$$= \max_{p_{m_\mathcal{M}}(\cdot)} \max_{\substack{P_i(m), \forall i \\ s.t.\, \mathbb{E}[P_i(m)] \leq P}} \max_{\substack{p_{x_{s,\mathcal{M}}|m_\mathcal{M}}(\cdot)\, s.t. \\ \mathbb{E}[||x_i||^2 | m_\mathcal{M}=m] \leq P_i(m), \forall i, \forall m}} \min_{\Omega} I(x_\Omega; y_{\Omega^c}|m_\mathcal{M}, x_{\Omega^c}) + N$$

$$\leq \max_{p_{m_\mathcal{M}}(\cdot)} \min_{\Omega} \max_{\substack{P_i(m), \forall i \\ s.t.\, \mathbb{E}[P_i(m)] \leq P}} \max_{\substack{p_{x_{s,\mathcal{M}}|m_\mathcal{M}}(\cdot)\, s.t. \\ \mathbb{E}[||x_i||^2 | m_\mathcal{M}=m] \leq P_i(m), \forall i, \forall m}} I(x_\Omega; y_{\Omega^c}|m_\mathcal{M}, x_{\Omega^c}) + N$$

$$= \max_{\substack{t_m \geq 0,\, s.t. \\ \sum_m t_m = 1}} \min_{\Omega} \max_{\substack{P_i(m), \forall i \\ s.t.\, \mathbb{E}[P_i(m)] \leq P}} \max_{\substack{p_{x_{s,\mathcal{M}}|m_\mathcal{M}}(\cdot)\, s.t. \\ \mathbb{E}[||x_i||^2 | m_\mathcal{M}=m] \leq P_i(m), \forall i, \forall m}} \sum_m t_m I(x_\Omega; y_{\Omega^c}|m_\mathcal{M}=m, x_{\Omega^c}) + N, \quad (51)$$

where we use $m$ to enumerate the $2^N$ states of the network and to simplify notation $t_m = p_{m_\mathcal{M}}(m)$. Clearly, the inner most maximization in the above expression leads to Gaussian $p_{x_{s,\mathcal{M}}|m_\mathcal{M}}(\cdot|m)$ for each state $m$ with the variance of $x_i$ at state $m$ equal to $P_i(m)$. Therefore the inner most maximization reduces to optimizing the covariance matrix of $x_\Omega$ for each state $m$ under the constraint that the diagonal entry of this matrix corresponding to $i \in \Omega$ should be smaller than $P_i(m)$.

We will next argue that if we consider independent transmissions from the nodes in the network, corresponding to an identity covariance matrix, and discard the optimization of the power allocation across the states $m$, i.e., take $P_i(m) = P$, $\forall i$, and $\forall m$, the gap to the expression in (51) is upper bounded by $2N$, which leads to the conclusion that

$$\overline{C}_{h,d} \leq \max_{t_m \geq 0,\, s.t.\, \sum_m t_m = 1} \min_{\Omega} \sum_m t_m I(x_\Omega^m; y_{\Omega^c}^m | x_{\Omega^c}^m) + 3N, \quad (52)$$

where $\{x_i, i \in \{s, \mathcal{M}\} \text{ and } m_i = T\}$ are independent, each with distribution $\mathcal{CN}(0, P)$. $x_\Omega^m = \{x_i, i \in \Omega \text{ and } m_i = T\}$, $y_{\Omega^c}^m = \{y_i, i \in \Omega \text{ and } m_i = L\}$ and $x_{\Omega^c}^m = \{x_i, i \in \Omega^c \text{ and } m_i = T\}$.

To prove (52), in the sequel we consider a MIMO channel with $N_R$ receive and $N_L$ transmit antennas, $N_R \times N_L$ channel matrix $H$ and a total average transmit power constraint of $N_L P$ at the transmitter. Let us assume that there are a number of states for communicating over this channel, state $m$ occurring with probability $t_m$ and $\sum_m t_m = 1$, where each state corresponds to using a subset of the transmit and receive antennas. In other words, each state induces a sub-MIMO channel with a channel matrix $H^m$ that contains a subset of the rows and the columns of the original channel matrix $H$. Let $\sigma_{i,m}$ denote the singular values of the matrix $H^m$, some of which can be zero. We next prove that

$$\max_{\substack{P_i(m),\\ s.t. \sum_{m,i} t_m P_i(m) \leq N_L P}} \sum_m t_m \sum_{i=1}^K \log(1+\sigma_{i,m}^2 P_i(m)) - \sum_m t_m \sum_{i=1}^K \log\left(1+\sigma_{i,m}^2 P\right) \leq \frac{N_L}{e} + K, \qquad (53)$$

where $K = \min(N_R, N_L)$. Note that the difference between the two terms above upperbounds the difference between the first term in (51) and the first term in (52) because the mutual information terms in (51) and (52) correspond to a MIMO channel between $\mathbf{x}_\Omega$ and $\mathbf{y}_{\Omega^c}$. In (51), optimal power allocation across the eigenvalues of the channel matrices induced at different states is allowed, while in (52) we allocate equal power to all eigenvalues at all states.

We will prove that the upper bound in (53) on the difference of the two terms holds for any schedule $\{t_m\}$ and any power allocation strategy $\{P_i(m)\}$. For any $\{t_m\}$ and $\{P_i(m)\}$, we have

$$\sum_{i=1}^K \sum_m t_m \log\left(\frac{1+\sigma_{i,m}^2 P_i(m)}{1+\sigma_{i,m}^2 P}\right)$$

$$\stackrel{(a)}{\leq} \sum_{i=1}^K \sum_m t_m \log\left(\frac{1+\sigma_{i,m}^2 P_i(m)}{\max\{1,\sigma_{i,m}^2 P\}}\right) \qquad (54)$$

$$= \sum_{i=1}^K \sum_m t_m \log\left(\frac{1}{\max\{1,\sigma_{i,m}^2 P\}} + \frac{\sigma_{i,m}^2 P_i(m)}{\max\{1,\sigma_{i,m}^2 P\}}\right), \qquad (55)$$

$$\leq \sum_{i=1}^K \sum_m t_m \log\left(1 + \frac{\sigma_{i,m}^2 P_i(m)}{\sigma_{i,m}^2 P}\right), \qquad (56)$$

where $(a)$ follows by lower bounding the denominator of the first term, $1+\sigma_{i,m}^2 P$ by $\max\{1,\sigma_{i,m}^2 P\}$. Now, we will use Jensen's inequality to further bound (56), as follows

$$\sum_{i=1}^K \sum_m t_m \log\left(1+\frac{P_i(m)}{P}\right) \leq \sum_{i=1}^K \log\left(1+\sum_m t_m \frac{P_i(m)}{P}\right) = \sum_{i=1}^K \log\left(1+\frac{P_i}{P}\right)$$

where we define $P_i = \sum_m t_m P_i(m)$. Now, we use the fact that $\sum_{i=1}^K P_i \leq N_L P$ to see that, due to the waterfilling solution,

$$\sum_{i=1}^{K)} \log\left(1+\frac{P_i}{P}\right) \leq \sum_{i=1}^K \log\left(1+\frac{N_L}{K}\right) = K\log\left(1+\frac{N_L}{K}\right)$$

$$= K\log\left(\frac{N_L}{K}\right) + K\log\left(1+\frac{K}{N_L}\right)$$

$$\stackrel{(d)}{\leq} \underbrace{K\log\left(\frac{N_L}{K}\right)}_{\log\left(\frac{N_L}{K}\right)^K} + K \stackrel{(c)}{\leq} \frac{N_L}{e} + K,$$

where (d) follows because $K \leq N_L$ and (c) follows because $\max_K \left(\frac{N_L}{K}\right)^K \leq e^{N_L/e}$ and we also take natural logarithms. For simplifying the statement of the result, in Theorem 2.3, we just upper bound $\frac{N_L}{e} + K \leq 2N_L$ and note that for the MIMO channel induced by any cut $\Omega$ the number of transmit antennas are smaller than $N$, the total number of nodes in the network.

## D. QMF in Half-Duplex Networks

The main result of the earlier section in (52) shows that i.i.d. Gaussian distribution along with *uniform* power allocation and a *fixed* schedule is within an additive constant of $N + 2\sum_i M_i$ of the information-theoretic cutset upper bound on the capacity of half-duplex networks. It is straightforward to argue that the rate of any fixed schedule under i.i.d. Gaussian distributions and *uniform* power allocation can be approximately achieved using a QMF strategy. This was already demonstrated in Theorem 8.3 of [1] and here we briefly summarize the main idea.

Fix a schedule $t_m, m = 1, \ldots, 2^N$ s.t. $\sum_m t_m = 1$ for the half-duplex network. Divide the total bandwidth of the network to $2^N$ bands of width $t_i W$, for $i = 1, \ldots, 2^N$. Each mode of the network operates over the corresponding band $t_i W$ and therefore the half-duplex constraint is satisfied, no node transmits and receives simultaneously over the same frequency band. Different frequency bands can be thought as a MIMO channel with a diagonal channel transfer matrix. Therefore the lattice QMF strategy developed in Section IV for multiple antenna networks can be applied to this setup and by Theorem 2.1 will achieve a rate in bits/s

$$R \geq \min_{\Omega} \sum_m t_m W I(x_\Omega^m; y_{\Omega^c}^m | x_{\Omega^c}^m) - 2WN$$

where $x_i^m, i \in \{s, \mathcal{M}\}, m = 1, \ldots, 2^N$ are i.i.d. Gaussian $\mathcal{CN}(0, t_m P)$ if node $i$ is transmitting in state $m$ and $x_\Omega^m = \{x_i^m, i \in \Omega\}$. [14] Choosing the fixed schedules $t_m, m = 1, \ldots 2^N$ that maximizes the above rate, we observe that we can achieve the right-hand side of (52) within $5N$ bits/s/Hz. The result in Theorem 2.3 is a straightforward generalization of the above arguments to the case when nodes contain multiple transmit and receive antennas.

## References

[1] S. Avestimehr, S N. Diggavi and D. Tse, *Wireless network information flow: a deterministic approach*, IEEE Transactions on Information theory, Vol. 57, No. 4, pp 1872–1905, April 2011. See also eprint arXiv:0906.5394v2 - arxiv.org.

[2] M. Anand and P. R. Kumar, *A digital interface for Gaussian relay and interference networks: Lifting codes from the discrete superposition model,* IEEE Transactions on Information Theory: Special Issue on Interference Networks, May 2011.

[3] A. Raja and P. Viswanath, "Compress-and-Forward Scheme for a Relay Network: Approximate Optimality and Connection to Algebraic Flows", preprint http://arxiv.org/abs/1012.0416, January 2011.

[4] A. Ozgur and S N. Diggavi. Approximately achieving Gaussian relay network capacity with lattice codes. *Proc. of IEEE ISIT 2010, Austin, Texas*, pp 669–673, June 2010. See also http://arxiv.org/abs/1005.1284.

[5] S. Lim, Y.-H. Kim, A. El-Gamal, and S-Y.Chung, "Noisy network coding," *IEEE Transactions on Information Theory*, vol. 57, no. 5, p. 31323152, May 2011.

[6] T. M. Cover and A. El Gamal, *Capacity theorems for the relay channel*, IEEE Trans. on Information Theory, vol.25, no.5, pp.572-584, September 1979.

[7] G. Kramer and J. Hou, "On message lengths for noisy network coding," in *ITW*, October 2011, pp. 430–431.

[8] T. M. Cover and J. A. Thomas, *Elements of Information Theory*, Wiley & Sons Inc., 1991.

---

[14] Note that since the noise accumulated over bandwidth $t_m W$ has variance $t_m N_0 W$, the spectral efficiency achieved with power $t_m P$ over bandwidth $t_m W$ is equal to the one with power $P$ over bandwidth $W$.

[9] G. Kramer, I. Maric, and R. Yates, *Cooperative Communications*. Foundations and Trends in Networking, 2006.

[10] E. Perron, *Information-theoretic secrecy for wireless networks,* Ph.D. thesis, EPFL Lausanne, 2009

[11] S. H. Lim, Y.-H. Kim, A. El Gamal, and S.-Y. Chung, Noisy Network Coding, submitted to IEEE Transactions on Information Theory, Feb. 2010. arXiv:1002.3188v1.

[12] G. Kramer, *Models and Theory for Relay Channels with Receive Constraints,* in Proc. 42th Annual Allerton Conference, Sept.2004.

[13] M. Yuksel and E. Erkip, *Multi-antenna cooperative wireless systems: A diversity multiplexing tradeoff perspective,* IEEE Trans. Information Theory, vol. 53, pp. 33713393, Oct, 2007.

[14] W. Nam and S.-Y. Chung, *Relay networks with orthogonal components*, in Proc. 46th Annual Allerton Conference, Sept.2008.

[15] U. Erez and R. Zamir, *Achieving $\frac{1}{2}\log(1+SNR)$ on the AWGN channel with lattice encoding and decoding,* IEEE Trans. on Information Theory, vol.50, no.10, pp.2293-2514, Oct. 2004.

[16] U. Erez, S. Litsyn and R. Zamir, *Lattices which are good for (almost) everything,* IEEE Trans. on Information Theory, vol.51, no.10, pp.3401-3416, Oct. 2005.

[17] J. H. Conway and N. J. A. Sloane, *Sphere Packings, Lattices and Groups*, Springer-Verlang, New York, 1998.

[18] H.-A. Loeliger, *Averaging bounds for lattices and linear codes*, IEEE Trans. on Information theory, vol.43, pp. 1767-1773, Nov. 1997.

[19] D. Krithivasan and S. Pradhan, *A proof of the existence of good lattices,* tech. rep., University of Michigan, July 2007. See http://www.eecs.umich.edu/techreports/systems/cspl/cspl-384.pdf.

## VII. Appendix

We first introduce the following two technical lemmas that we use repeatedly in this appendix.

*Lemma 7.1: (Lemma 11 of [15])*

(a) Let $\mathbf{u} \sim \text{unif}(\mathcal{B}(R))$. Let us denote $\frac{1}{n}\mathbb{E}[\|\mathbf{u}\|^2] = \frac{R^2}{n+2} := \sigma^2$. Let $\mathbf{z} \sim \mathcal{N}(0, \sigma^2 I_n)$. Then,

$$f_{\mathbf{u}}(\mathbf{x}) \leq f_{\mathbf{z}}(\mathbf{x}) \, e^{n\epsilon_2},$$

where $\epsilon_2 = \frac{1}{2}\log(2\pi e G_n^*) + \frac{1}{n}$.

(b) Let $\mathbf{u} \sim \text{unif}(\mathcal{V})$ where $\mathcal{V}$ is the Voronoi region of a lattice $\Lambda$. Note that $\frac{1}{n}\mathbb{E}[\|\mathbf{u}\|^2] = \sigma^2(\Lambda)$. Let $\mathbf{z} \sim \mathcal{N}(0, \sigma^2 I_n)$ such that

$$\sigma^2 = \frac{G_n^*}{G(\Lambda)}(\rho_{cov}(\Lambda))^2 \sigma^2(\Lambda).$$

Then,

$$f_{\mathbf{u}}(\mathbf{x}) \leq f_{\mathbf{z}}(\mathbf{x}) \, e^{n\epsilon_2(\Lambda)},$$

where $\epsilon_2(\Lambda) = \log(\rho_{cov}(\Lambda)) + \frac{1}{2}\log(2\pi e G_n^*) + \frac{1}{n}$.

The significance of the above lemma is that it allows to upper bound the probability distribution of a random variable $\mathbf{u}$, either uniformly distributed on an n-dimensional sphere or over the Voronoi region of a Rogers-good lattice, with the probability distribution of a Gaussian vector of identity covariance matrix and of the same variance with $\mathbf{u}$. Note that $\epsilon_2$ in part (a) of the lemma goes to zero with increasing dimension $n$. Similarly in part (b), $\epsilon_2(\Lambda) \to 0$ and $\sigma^2 \to \sigma^2(\Lambda)$ as $n$ increases if $\Lambda$ is Rogers-good.

*Lemma 7.2:* Let $z_i, i = 1, \ldots, n$ be independent random variables with distribution $\mathcal{N}(0, \gamma_i^2)$. Then,

$$\mathbb{P}\left(\sum_{i=1}^{n} z_i^2 \leq nc\right) \leq e^{-\left(\frac{1}{2}\sum_{i=1}^{n}\log(1+2\gamma_i^2 t) - ntc\right)}$$

for any $t > 0$. When $\gamma_i^2 = \gamma^2$, $\forall i$, such that $\gamma^2 > c$, we have

$$\mathbb{P}\left(\sum_{i=1}^n z_i^2 \leq nc\right) \leq e^{-\frac{n}{2}\left(\log\left(\frac{\gamma^2}{c}\right) - 1 + \frac{c}{\sigma^2}\right)}.$$

*Proof of Lemma 7.2:* The proof of the lemma follows by a simple application of the exponential Chebyshev's inequality. For any $t > 0$, we have

$$\mathbb{P}\left(\sum_{i=1}^n z_i^2 \leq nc\right) = \mathbb{P}\left(e^{-t\sum_{i=1}^n z_i^2} \geq e^{-ntc}\right) \leq \mathbb{E}[e^{-t\sum_{i=1}^n z_i^2}]\, e^{ntc} = \prod_{i=1}^n \mathbb{E}[e^{-tz_i^2}]\, e^{ntc}$$

$$= \prod_{i=1}^n \left(\frac{1}{\sqrt{1+2\gamma_i^2 t}}\right) e^{ntc} = e^{-\left(\frac{1}{2}\sum_{i=1}^n \log(1+2\gamma_i^2 t) - ntc\right)}.$$

When $\gamma_i^2 = \gamma^2$, $\forall i$, choosing $t = (\frac{1}{2c} - \frac{1}{2\gamma^2})$ yields

$$\mathbb{P}\left(\sum_{i=1}^n z_i^2 \leq nc\right) \leq e^{-\sup_{t \geq 0}\left(\frac{n}{2}\log(1+2\gamma^2 t) - ntc\right)} \leq e^{-\frac{n}{2}\left(\log\left(\frac{\gamma^2}{c}\right) - 1 + \frac{c}{\sigma^2}\right)}.$$

$\square$

The proof of Lemma 5.4 follows by a straightforward application of the above two lemmas.

*Proof of Lemma 5.4:* If $\nu$ is uniformly distributed over $\mathcal{V}^Q$, by part-(b) of Lemma 7.1 we have

$$\mathbb{P}\left(\|\nu\|^2 \leq n\,\sigma_c^2\right) \leq e^{n\epsilon_2(\Lambda^Q)} \mathbb{P}\left(\|\nu'\|^2 \leq n\,\sigma_c^2\right)$$

where $\nu' \sim \mathcal{N}(0, \sigma_\nu^2 I_n)$ with

$$\sigma_\nu^2 = \frac{G_n^*}{G(\Lambda^Q)} (\rho_{cov}(\Lambda^Q))^2\, \sigma^2(\Lambda^Q) = (1 + o_n(1))\sigma^2(\Lambda^Q).$$

Applying Lemma 7.2 for the case of equal variances yields the result

$$\mathbb{P}\left(\|\nu'\|^2 \leq n\,\sigma_c^2\right) \leq e^{-\frac{n}{2}\left(\log\left(\frac{(1+o_n(1))\sigma^2(\Lambda^Q)}{\sigma_c^2}\right) - 1 + \frac{\sigma_c^2}{(1+o_n(1))\sigma^2(\Lambda^Q)}\right)},$$

and therefore

$$\mathbb{P}\left(\|\nu\|^2 \leq n\,\sigma_c^2\right) \leq e^{n\epsilon_2(\Lambda^Q)} e^{-\frac{n}{2}\left(\log\left(\frac{(1+o_n(1))\sigma^2(\Lambda^Q)}{\sigma_c^2}\right) - 1 + \frac{\sigma_c^2}{(1+o_n(1))\sigma^2(\Lambda^Q)}\right)} = e^{-\frac{n}{2}\left(\log\left(\frac{\sigma^2(\Lambda^Q)}{\sigma_c^2}\right) - 1 + \frac{\sigma_c^2}{\sigma^2(\Lambda^Q)} - o_n(1)\right)}.$$

$\square$

To prove Lemmas 5.1 and 5.2, we introduce the following lemma as an intermediate step:

*Lemma 7.3:* Let $\mathbf{x}_j, \mathbf{x}'_j$, $j = 1, \ldots, N_1$ be independent discrete random variables uniformly distributed over the $p_r^n$ lattice points $p_r^{-1} \Lambda^T \cap \mathcal{V}^T$. Let $\mathbf{z}_i$ and $\mathbf{u}'_i$, $i = 1, \ldots, N_2$ be independent random variables with distributions $\mathbf{z}_i \sim \mathcal{N}(0, \sigma^2 I_n)$ and $\mathbf{u}'_i \sim \text{unif}(\mathcal{V}_{1,i}^Q)$ where $\mathcal{V}_{1,i}^Q$ denotes the Voronoi region of the lattice $\Lambda_{1,i}^Q$. Let $\mathcal{S}_1, \ldots, \mathcal{S}_{N_2} \subseteq \mathbb{R}^n$. Then,

$$\mathbb{P}\left(\sum_{j=1}^{N_1} h_{ij}(\mathbf{x}_j - \mathbf{x}'_j) + \mathbf{z}_i - \mathbf{u}'_i \in \mathcal{S}_i,\; \forall i = 1, \ldots, N_2\right)$$

$$\leq \left((1 + \epsilon_4(\Lambda^T))^{N_2} e^{n\epsilon_1(\Lambda^T) + n\epsilon_2}\right)^{2N_1} \left(e^{n\epsilon_2(\Lambda_{1,i}^Q)}\right)^{N_2} \mathbb{P}\left(\sum_{j=1}^{N_1} h_{ij}(\tilde{\mathbf{x}}_j - \tilde{\mathbf{x}}'_j) + \tilde{\mathbf{z}}_i \in \mathcal{S}_i,\; \forall i = 1, \ldots, N_2\right)$$

where $\tilde{\mathbf{x}}_j, \tilde{\mathbf{x}}'_j, j = 1, \ldots, N_1, \tilde{\mathbf{z}}_i, i = 1, \ldots, N_2$ are all independent Gaussian random variables such that $\tilde{\mathbf{x}}_j, \tilde{\mathbf{x}}'_j \sim \mathcal{N}(0, \sigma_x^2 I_n)$ with

$$\sigma_x^2 = (1 + p_r^{-1})^2 (\rho_{cov}(\Lambda^T))^2 \frac{G_n^*}{G(\Lambda^T)} \sigma^2(\Lambda^T)$$

and $\tilde{\mathbf{z}}_i \sim \mathcal{N}(0, \sigma_z^2 I_n)$,

$$\sigma_z^2 = (1 + \epsilon_5)^{2N_1} \left(1 + \frac{G_n^*}{G(\Lambda_{1,i}^Q)} (\rho_{cov}(\Lambda_{1,i}^Q))^2 \sigma^2(\Lambda_{1,i}^Q)\right)$$

where all $\epsilon_1(\Lambda^T), \epsilon_2, \epsilon_2(\Lambda_{1,i}^Q) \epsilon_4(\Lambda^T), \epsilon_5 \to 0$ as $n \to \infty$. Furthermore $\sigma_x^2 \to \sigma^2(\Lambda^T)$ and $\sigma_z^2 \to 1 + \sigma^2(\Lambda_{1,i}^Q)$ since both $\Lambda^T$ and $\Lambda_{1,i}^Q$ are Rogers-good.

*Proof of Lemma 7.3:* First, by using Part-(b) of Lemma 7.1, we can upper bound the probability

$$\mathbb{P}\left(\sum_{j=1}^{N_1} h_{ij}(\mathbf{x}_j - \mathbf{x}'_j) + \mathbf{z}_i - \mathbf{u}'_i \in \mathcal{S}_i, \forall i = 1, \ldots, N_2\right)$$

by

$$\left(e^{n\epsilon_2(\Lambda_{1,i}^Q)}\right)^{N_2} \mathbb{P}\left(\sum_{j=1}^{N_1} h_{ij}(\mathbf{x}_j - \mathbf{x}'_j) + \mathbf{z}_{eq,i} \in \mathcal{S}_i, \forall i = 1, \ldots, N_2\right), \tag{57}$$

where $\mathbf{z}_{eq,i}$ are i.i.d with distribution $\mathcal{N}(0, \sigma_{eq}^2 I_n)$,

$$\sigma_{eq}^2 = 1 + \frac{G_n^*}{G(\Lambda_{1,i}^Q)} (\rho_{cov}(\Lambda_{1,i}^Q))^2 \sigma^2(\Lambda_{1,i}^Q).$$

Since $\Lambda_{1,i}^Q$ is Rogers-good, $\epsilon_2(\Lambda_{1,i}^Q)$ given in the lemma vanishes with increasing $n$. The probability in (57) can be expressed as,

$$\mathbb{P}\left(\sum_{j=1}^{N_1} h_{ij}(\mathbf{x}_j - \mathbf{x}'_j) + \mathbf{z}_{eq,i} \in \mathcal{S}_i, \forall i = 1, \ldots, N_2\right)$$

$$= \left(p_r^{-n}\right)^{2N_1} \sum_{\substack{\mathbf{x}_1, \ldots, \mathbf{x}_{N_1}, \mathbf{x}'_1, \ldots, \mathbf{x}'_{N_1} \\ \in p_r^{-1}\Lambda^T \cap \mathcal{V}^T}} \mathbb{P}\left(\sum_{j=1}^{N_1} h_{ij}(\mathbf{x}_j - \mathbf{x}'_j) + \mathbf{z}_{eq,i} \in \mathcal{S}_i, \forall i = 1, \ldots, N_2\right). \tag{58}$$

The last probability is only over $\mathbf{z}_{eq,i}$'s and note that the $\mathbf{x}_j$ and $\mathbf{x}'_j$'s now denote the dummy variables of the summation. Consider one of the summations above of the form,

$$p_r^{-n} |\mathcal{V}^T| \sum_{\mathbf{x}_1 \in p_r^{-1}\Lambda^T \cap \mathcal{V}^T} \mathbb{P}\left(\sum_{j=1}^{N_1} h_{ij}(\mathbf{x}_j - \mathbf{x}'_j) + \mathbf{z}_{eq,i} \in \mathcal{S}_i, \forall i = 1, \ldots, N_2\right),$$

where $\mathbf{x}_1$ denotes the dummy variable of the summation and $\mathbf{x}_2, \ldots, \mathbf{x}_{N_1}, \mathbf{x}'_1, \ldots, \mathbf{x}'_{N_1}$ are fixed vectors. We show below that this summation is upper bounded by

$$(1 + \epsilon_4(\Lambda^T))^{N_2} \int_{\mathcal{V}^T + p_r^{-1}\mathcal{V}^T} d\mathbf{x}_1 \, \mathbb{P}\left(\sum_{j=1}^{N_1} h_{ij}(\mathbf{x}_j - \mathbf{x}'_j) + \mathbf{z}'_{eq,i} \in \mathcal{S}_i, \forall i = 1, \ldots, N_2\right) \tag{59}$$

where $\mathbf{z}'_{eq,i} \sim \mathcal{N}(0, (1+\epsilon_5)\sigma_{eq}^2 I_n)$ and both $\epsilon_4(\Lambda)$ and $\epsilon_5 \to 0$ as $n \to 0$. For two sets $A \subset \mathbb{R}^n$ and $B \subset \mathbb{R}^n$, the sum set $A + B \subset \mathbb{R}^n$ denotes $A + B = \{\mathbf{a} + \mathbf{b} : \mathbf{a} \in A, \mathbf{b} \in B\}$. Applying this upper bound recursively to all the summations in (58) yields

$$\mathbb{P}\left(\sum_{j=1}^{N_1} h_{ij}(\mathbf{x}_j - \mathbf{x}'_j) + \mathbf{z}_{eq,i} \in \mathcal{S}_i, \ \forall i = 1, \ldots, N_2\right)$$

$$\leq (1 + \epsilon_4(\Lambda^T))^{2N_2 N_1} \frac{1}{|\mathcal{V}^T|^{2N_1}} \int_{\mathcal{V}^T + p_r^{-1}\mathcal{V}^T} \cdots \int_{\mathcal{V}^T + p_r^{-1}\mathcal{V}^T} d\mathbf{x}_1 \ldots d\mathbf{x}_{N_1} d\mathbf{x}'_1 \ldots d\mathbf{x}'_{N_1}$$

$$\mathbb{P}\left(\sum_{j=1}^{N_1} h_{ij}(\mathbf{x}_j - \mathbf{x}'_j) + \tilde{\mathbf{z}}_i \in \mathcal{S}_i, \ \forall i = 1, \ldots, N_2\right)$$

$$\leq (1 + \epsilon_4(\Lambda^T))^{2N_2 N_1} \frac{1}{|\mathcal{V}^T|^{2N_1}} \int_{\mathcal{B}\left((1+p_r^{-1})R_u^T\right)} \cdots \int_{\mathcal{B}\left((1+p_r^{-1})R_u^T\right)} d\mathbf{x}_1 \ldots d\mathbf{x}_{N_1} d\mathbf{x}'_1 \ldots d\mathbf{x}'_{N_1}$$

$$\mathbb{P}\left(\sum_{j=1}^{N_1} h_{ij}(\mathbf{x}_j - \mathbf{x}'_j) + \tilde{\mathbf{z}}_i \in \mathcal{S}_i, \ \forall i = 1, \ldots, N_2\right) \tag{60}$$

where $\tilde{\mathbf{z}}_i \sim \mathcal{N}(0, (1+\epsilon_5)^{2N_1} \sigma_{eq}^2 I_n)$. $R_u^T$ in the last inequality denotes the covering radius of $\mathcal{V}^T$ and $\mathcal{B}\left((1 + p_r^{-1})R_u^T\right)$ denotes an $n$-dimensional sphere in $\mathbb{R}^n$ of radius $(1 + p_r^{-1})R_u^T$. The last inequality follow follows the fact that $\mathcal{V}^T + p_r^{-1}\mathcal{V}^T \subseteq \mathcal{B}\left((1 + p_r^{-1})R_u^T\right)$ which in turn follows from the definition of $R_u^T$. We can rewrite (60) as

$$(1 + \epsilon_4(\Lambda^T))^{2N_2 N_1} \left(e^{n\epsilon_1(\Lambda^T)}\right)^{2N_1} \frac{1}{\left|\mathcal{B}\left((1+p_r^{-1})R_u^T\right)\right|^{2N_1}} \int_{\mathcal{B}\left((1+p_r^{-1})R_u^T\right)} \cdots \int_{\mathcal{B}\left((1+p_r^{-1})R_u^T\right)}$$

$$d\mathbf{x}_1 \cdots d\mathbf{x}_{N_1} d\mathbf{x}'_1 \cdots d\mathbf{x}'_{N_1} \mathbb{P}\left(\sum_{j=1}^{N_1} h_{ij}(\mathbf{x}_j - \mathbf{x}'_j) + \tilde{\mathbf{z}}_i \in \mathcal{S}_i, \ \forall i = 1, \ldots, N_2\right) \tag{61}$$

where,

$$\frac{\left|\mathcal{B}\left((1+p_r^{-1})R_u^T\right)\right|}{|\mathcal{V}^T|} = \frac{\left|\mathcal{B}\left((1+p_r^{-1})R_u^T\right)\right|}{\left|\mathcal{B}\left(R_l^T\right)\right|} = \left(\frac{(1+p_r^{-1})R_u^T}{R_l^T}\right)^n = e^{n\epsilon_1(\Lambda^T)}$$

and $\epsilon_1(\Lambda^T) = \log(1 + p_r^{-1}) + \log \rho_{cov}(\Lambda^T)$. Recall that the effective radius $R_l^T$ of the lattice $\Lambda^T$ is defined as the radius of a sphere having the same volume as the Voronoi region of $\Lambda^T$. Since $\Lambda^T$ is Rogers-good and $p_r \to \infty$ as $n \to \infty$, we have $\epsilon_1(\Lambda^T) \to 0$. We can upper bound (61) by applying Part-(a) of Lemma 7.1 which gives

$$\mathbb{P}\left(\sum_{j=1}^{N_1} h_{ij}(\mathbf{x}_j - \mathbf{x}'_j) + \mathbf{z}_{eq,i} \in \mathcal{S}_i, \ \forall i = 1, \ldots, N_2\right)$$

$$\leq \left((1 + \epsilon_4(\Lambda^T))^{N_2} e^{n\epsilon_1(\Lambda^T) + n\epsilon_2}\right)^{2N_1} \mathbb{P}\left(\sum_{j=1}^{N_1} h_{ij}(\tilde{\mathbf{x}}_j - \tilde{\mathbf{x}}'_j) + \tilde{\mathbf{z}}_i \in \mathcal{S}_i, \ \forall i = 1, \ldots, N_2\right), \tag{62}$$

where $\tilde{\mathbf{x}}_j, \tilde{\mathbf{x}}'_j, j = 1, \ldots, N_1$ are independent $\sim \mathcal{N}(0, \sigma_x^2 I_n)$ with

$$\sigma_x^2 = \frac{\left((1+p_r^{-1})R_u^T\right)^2}{n+2}. \tag{63}$$

Plugging the expression in (16) to (63), yields

$$\sigma_x^2 = \frac{\left((1+p_r^{-1})R_u^T\right)^2}{n+2} = (1+p_r^{-1})^2 (\rho_{cov}(\Lambda^T))^2 \frac{G_n^*}{G(\Lambda^T)} \sigma^2(\Lambda^T).$$

The upper bounds (57) and (62) together yield the result stated in the lemma.

It remains to prove (59). We will first show that

$$p_r^{-n}|\mathcal{V}^T| \, \mathbb{P}\left(\sum_{j=1}^{N_1} h_{ij}(\mathbf{x}_j - \mathbf{x}'_j) + \mathbf{z}_{eq,i} \in \mathcal{S}_i, \, \forall i = 1, \ldots, N_2\right)$$

$$\leq (1 + \epsilon_4(\Lambda^T))^{N_2} \int_{p_r^{-1}\mathcal{V}^T} d\mathbf{s} \, \mathbb{P}\left(h_{i1}\mathbf{s} + \sum_{j=1}^{N_1} h_{ij}(\mathbf{x}_j - \mathbf{x}'_j) + \mathbf{z}'_{eq,i} \in \mathcal{S}_i, \, \forall i = 1, \ldots, N_2\right) \quad (64)$$

where $\mathbf{x}_j$, $\mathbf{x}'_j$'s are fixed vectors and $\mathbf{z}_{eq,i} \sim \mathcal{N}(0, \sigma_{eq}^2 I_n)$, $\mathbf{z}'_{eq,i} \sim \mathcal{N}(0, (1+\epsilon_5)\sigma_{eq}^2 I_n)$ and both $\epsilon_4(\Lambda)$ and $\epsilon_5 \to 0$ as $n \to 0$.

First, note that for $\mathbf{z}'_{eq,i} \sim \mathcal{N}(0, \delta^2 I_n)$, $i = 1, \ldots, N_2$,

$$\mathbb{P}\left(h_{i1}\mathbf{s} + \sum_{j=1}^{N_1} h_{ij}(\mathbf{x}_j - \mathbf{x}'_j) + \mathbf{z}'_{eq,i} \in \mathcal{S}_i, \, \forall i = 1, \ldots, N_2\right)$$

$$= \prod_{i=1}^{N_2} \mathbb{P}\left(h_{i1}\mathbf{s} + \sum_{j=1}^{N_1} h_{ij}(\mathbf{x}_j - \mathbf{x}'_j) + \mathbf{z}'_{eq,i} \in \mathcal{S}_i,\right)$$

$$= \prod_{i=1}^{N_2} \int_{\mathcal{S}_i} f_{\mathbf{z}'_{eq,i}}\left(\mathbf{z}_i - h_{i1}\mathbf{s} - \sum_{j=1}^{N_1} h_{ij}(\mathbf{x}_j - \mathbf{x}'_j)\right) d\mathbf{z}_i. \quad (65)$$

The probability density function $f_{\mathbf{z}'_{eq,i}}(\mathbf{c})$ of $\mathbf{z}'_{eq,i}$ depends only on $\|\mathbf{c}\|$. By the triangle inequality, for any two vectors $\mathbf{a}$ and $\mathbf{b}$, we have

$$\|\mathbf{a} + \mathbf{b}\|^2 \leq \|\mathbf{a}\|^2 + 2\|\mathbf{a}\|\|\mathbf{b}\| + \|\mathbf{b}\|^2.$$

Also for any $t > 0$,

$$\|\mathbf{a}\|\|\mathbf{b}\| \leq \frac{\|\mathbf{a}\|^2}{t} + t\|\mathbf{b}\|^2.$$

Therefore, for any $t > 0$,

$$\|\mathbf{a} + \mathbf{b}\|^2 \leq \left(1 + \frac{2}{t}\right)\|\mathbf{a}\|^2 + (1 + 2t)\|\mathbf{b}\|^2.$$

Using this inequality, we obtain

$$f_{\mathbf{z}'_{eq,i}}(\mathbf{a} + \mathbf{b}) \propto e^{-\frac{\|\mathbf{a}+\mathbf{b}\|^2}{2\delta^2}} \geq e^{-(1+\frac{2}{t})\frac{\|\mathbf{a}\|^2}{2\delta^2}} \, e^{-\frac{(1+2t)\|\mathbf{b}\|^2}{2\delta^2}} \propto f_{\mathbf{z}_{eq,i}}(\mathbf{a}) \, e^{-\frac{(1+2t)\|\mathbf{b}\|^2}{2\delta^2}}$$

where $\mathbf{z}_{eq,i} \sim \mathcal{N}(0, \sigma_{eq}^2 I_n)$ with $\sigma_{eq}^2 = \left(1 + \frac{2}{t}\right)^{-1} \delta^2$. Applying this inequality to (65) with $\mathbf{a}_i = \mathbf{z}_i - \sum_{j=1}^{N_1} h_{ij}(\mathbf{x}_j - \mathbf{x}'_j)$ and $\mathbf{b}_i = h_{i1}\mathbf{s}$ yields

$$\mathbb{P}\left(h_{i1}\mathbf{s} + \sum_{j=1}^{N_1} h_{ij}(\mathbf{x}_j - \mathbf{x}'_j) + \mathbf{z}'_{eq,i} \in \mathcal{S}_i, \, \forall i = 1, \ldots, N_2\right)$$

$$\geq \prod_{i=1}^{N_2} \int_{\mathcal{S}_i} e^{-\frac{(1+2t)\|h_{i1}\mathbf{s}\|^2}{2\delta^2}} f_{\mathbf{z}_{eq,i}}\left(\mathbf{z}_i - \sum_{j=1}^{N_1} h_{ij}(\mathbf{x}_j - \mathbf{x}'_j)\right) d\mathbf{z}_i,$$

$$\geq e^{-\frac{(1+2t)N_2}{2\delta^2} p_r^{-1} R_u^T} \mathbb{P}\left(\sum_{j=1}^{N_1} h_{ij}(\mathbf{x}_j - \mathbf{x}'_j) + \mathbf{z}_{eq,i} \in \mathcal{S}_i, \, \forall i = 1, \ldots, N_2\right), \quad (66)$$

where we make of use of the inequality

$$\|h_{i1}\mathbf{s}\| = |h_{i1}|\|\mathbf{s}\| \leq |h_{i1}|p_r^{-1}R_u^T. \tag{67}$$

From (16) for $R_u^T$ and the choice for $p_r$ in (19), we know that $p_r^{-1}R_u^T = O(\sqrt{n}e^{-\frac{nR_r}{(\log n)^2}}) \to 0$ as $n \to 0$. We choose $t$ such that $t^{-1} \to 0$, while $t p_r^{-1} R_u^T \to 0$. For example, choose $t = n$. Integrating both sides of the inequality (66) with respect to $\mathbf{s}$ over the region $p_r^{-1}\mathcal{V}^T$, this yields the desired result in (64) where we denote $1 + \epsilon_4(\Lambda^T) = e^{\frac{(1+t)}{2\delta^2}p_r^{-1}R_u^T}$ and $1 + \epsilon_5 = \left(1 + \frac{2}{t}\right)$.

The conclusion in (59) follows by combining (64) with the following observation,

$$\sum_{\mathbf{x}_1 \in p_r^{-1}\Lambda^T \cap \mathcal{V}^T} \int_{p_r^{-1}\mathcal{V}^T} d\mathbf{s}\ \mathbb{P}\left(h_{i1}(\mathbf{x}_1 + \mathbf{s}) - h_{i1}\mathbf{x}_1' + \sum_{j=2}^{N_i} h_{ij}(\mathbf{x}_j - \mathbf{x}_j') + \mathbf{z}_{eq,i} \in \mathcal{S}_i,\ \forall i = 1,\ldots,N_2\right)$$

$$\leq \int_{\mathcal{V}^T + p_r^{-1}\mathcal{V}^T} d\mathbf{x}_1\ \mathbb{P}\left(h_{i1}\mathbf{x}_1 - h_{i1}\mathbf{x}_1' + \sum_{j=2}^{N_i} h_{ij}(\mathbf{x}_j - \mathbf{x}_j') + \mathbf{z}_{eq,i} \in \mathcal{S}_i,\ \forall i = 1,\ldots,N_2\right).$$

This observation simply follows from the fact that the summation and the integration in the first case, together correspond to integrating the function

$$\mathbb{P}\left(h_{i1}\mathbf{x}_1 - h_{i1}\mathbf{x}_1' + \sum_{j=2}^{N_i} h_{ij}(\mathbf{x}_j - \mathbf{x}_j') + \mathbf{z}_{eq,i} \in \mathcal{S}_i,\ \forall i = 1,\ldots,N_2\right)$$

over the sum region $p_r^{-1}\Lambda^T \cap \mathcal{V}^T + p_r^{-1}\mathcal{V}^T$ which lies inside the second region $\mathcal{V}^T + p_r^{-1}\mathcal{V}^T$. $\square$

*Proof of Lemma 5.1:* For a given $i \in \{\mathcal{M}, d\}$ and a set of indices $k_j, k_j',\ j = 1,\ldots N_i$ we first consider the probability

$$\mathbb{P}\left(\sum_{j=1}^{N_i} h_{ij}(\mathbf{x}_j^{(k_j)} - \mathbf{x}_j^{(k_j')}) + \mathbf{z}_i - \mathbf{u}_i' \notin \mathcal{V}^Q\right), \tag{68}$$

where $N_i$ denotes the number of nodes $j$ that have non-zero channel coefficients to node $i$. $\mathbf{x}_j^{(k_j)}$ and $\mathbf{x}_j^{(k_j')}$ are independent and uniformly distributed over the $p_r^n$ lattice points $p_r^{-1}\Lambda^T \cap \mathcal{V}^T$, $\mathbf{z}_i \sim \mathcal{N}(0,\sigma^2)$, and $\mathbf{u}_i' \sim \text{unif}(\mathcal{V}_{1,i}^Q)$. Note that we can immediately apply Lemma 7.3 by identifying $\mathcal{S}_1$ in the lemma as the complement of $\mathcal{V}^Q$, and switch from the discrete distribution over the lattice points $p_r^{-1}\Lambda^T \cap \mathcal{V}^T$ for $\mathbf{x}_j^{(k_j)}$ and $\mathbf{x}_j^{(k_j')}$ to a Gaussian distribution. More precisely the above probability is upper bounded by

$$\left((1 + \epsilon_4(\Lambda^T))\,e^{n\epsilon_1(\Lambda^T) + n\epsilon_2}\right)^{2N_i} e^{n\epsilon_2(\Lambda_{1,i}^Q)} \mathbb{P}\left(\sum_{j=1}^{N_i} h_{ij}(\tilde{\mathbf{x}}_j - \tilde{\mathbf{x}}_j') + \tilde{\mathbf{z}}_i \notin \mathcal{V}^Q\right),$$

where $\tilde{\mathbf{x}}_j, \tilde{\mathbf{x}}_j',\ j = 1,\ldots,N_i$ are independent $\sim \mathcal{N}(0, \sigma_x^2 I_n)$ with

$$\sigma_x^2 = (1 + p_r^{-1})^2 (\rho_{cov}(\Lambda^T))^2 \frac{G_n^*}{G(\Lambda^T)} \sigma^2(\Lambda^T),$$

and $\tilde{\mathbf{z}}_i \sim \mathcal{N}(0, \sigma_z^2 I_n)$,

$$\sigma_z^2 = (1 + \epsilon_5)^{2N_i}\left(1 + \frac{G_n^*}{G(\Lambda_{1,i}^Q)}(\rho_{cov}(\Lambda_{1,i}^Q))^2 \sigma^2(\Lambda_{1,i}^Q)\right)$$

where all $\epsilon_1(\Lambda^T),\ \epsilon_2,\ \epsilon_2(\Lambda_{1,i}^Q)\,\epsilon_4(\Lambda^T),\ \epsilon_5 \to 0$ as $n \to \infty$.

Note that $\sum_{j=1}^{N_i} h_{ij}(\tilde{\mathbf{x}}_j - \tilde{\mathbf{x}}'_j) + \tilde{\mathbf{z}}_i$ has distribution $\mathcal{N}(0, \sigma_i^2 I_n)$, where

$$\sigma_i^2 = 2 + 2 \sum_{j=1}^{N_i} |h_{ij}|^2 \, P + o_n(1),$$

which follows from our choices for $\sigma^2(\Lambda^Q_{1,i})$ and $\sigma^2(\Lambda^T)$ in (20) and (15) respectively. Note that both $\Lambda^T$ and $\Lambda^Q_{1,i}$ are Rogers-good and from (19), $p_r = e^{\frac{nR_r}{(\log n)^2}}$ and hence $p_r^{-1} \to 0$ as $n \to 0$. Since $\Lambda^Q$ is Poltyrev-good, we have

$$\mathbb{P}\left(\sum_{j=1}^{N_i} h_{ij}(\tilde{\mathbf{x}}_j - \tilde{\mathbf{x}}'_j) + \tilde{\mathbf{z}}_i \notin \mathcal{V}^Q\right) \leq e^{-n[E_P(\mu_i) - o_n(1)]} \tag{69}$$

where $E_P(\mu_i)$ is the Poltyrev exponent,

$$E_P(\mu_i) = \begin{cases} \frac{1}{2}[(\mu_i - 1) - \log \mu_i] & 1 < \mu_i \leq 2 \\ \frac{1}{2} \log \frac{e \mu_i}{4} & 2 \leq \mu_i \leq 4 \\ \frac{\mu_i}{8} & \mu_i \geq 4 \end{cases} \tag{70}$$

and $\mu_i = \sigma^2(\Lambda^Q)/\sigma_i^2$. By the union bound, for node $i \in \{\mathcal{M}, d\}$,

$$\mathbb{P}\left(\exists \{k_j, k'_j\} \text{ s.t. } \sum_{j=1}^{N_i} h_{ij}(\mathbf{x}_j^{(k_j)} - \mathbf{x}_j^{(k'_j)}) + \mathbf{z}_i - \mathbf{u}'_i \notin \mathcal{V}^Q\right)$$
$$\leq \left((1 + \epsilon_4(\Lambda^T))\, e^{n\epsilon_1(\Lambda^T) + n\epsilon_2}\right)^{2N_i} e^{n\epsilon_2(\Lambda^Q_{1,i})} \left(e^{2nR_r}\right)^{N_i} e^{-n[E_P(\mu_i) - o_n(1)]}$$

since for every $j = 1, \ldots, N_i$, $k_j$ and $k'_j$ run over the $e^{nR_r}$ possible transmit codewords. Finally,

$$\mathbb{P}\left(\exists i \in \{\mathcal{M}, d\}, \{k_j, k'_j\} \text{ s.t. } \sum_j h_{ij}(\mathbf{x}_j^{(k_j)} - \mathbf{x}_j^{(k'_j)}) + \mathbf{z}_i - \mathbf{u}'_i \notin \mathcal{V}^Q\right)$$
$$\leq \left((1 + \epsilon_4(\Lambda^T))\, e^{n\epsilon_1(\Lambda^T) + n\epsilon_2}\right)^{2N_i} e^{n\epsilon_2(\Lambda^Q_{1,i})} (N+1)\, e^{-n[E_P(\mu) - 2R_r N_s - o_n(1)]} \tag{71}$$

where $N_s = \max_{i \in \{\mathcal{M}, d\}} N_i$ and $\mu = \sigma^2(\Lambda^Q)/\sigma_s^2$ with

$$\sigma_s^2 = 2 + 2D_s + o_n(1).$$

Recall from (17) that $D_s = \max_{i \in \{\mathcal{M}, d\}} \sum_j |h_{ij}|^2 P$. We have chosen in (18) and (19)

$$R_r = \frac{1}{2} \log \sigma^2(\Lambda^Q) \qquad \text{and} \qquad \sigma^2(\Lambda^Q) = 2\eta(1 + D_s)$$

for some $\eta > 0$. Therefore $R_r$ increases logarithmically in $\eta$ while the Poltyrev exponent is linear in $\mu$ ( and hence in $\eta$) in the third regime in (70). By choosing the constant $\eta$ large enough, we can ensure that the exponent in (71) is negative and hence the probability decreases to zero when $n$ increases. $\square$

*Proof of Lemma 5.2:* Let us denote $N_L = |\Omega|$ and $N_R = |\Omega^c|$. We want to evaluate the probability

$$\mathbb{P}\left(\|\sum_{j \in \Omega_{l-1}} h_{ij}(\mathbf{x}_j^{(k_j)} - \mathbf{x}_j^{(k'_j)}) + \mathbf{z}_i - \mathbf{u}'_i\|^2 \leq n \sigma_c^2, \forall i \in \Omega^c\right).$$

where $\mathbf{x}_j^{(k_j)}$ and $\mathbf{x}_j^{(k'_j)}$, $j \in \Omega$ are independent and uniformly distributed over the $p_r^n$ lattice points $p_r^{-1}\Lambda^T \cap \mathcal{V}^T$, $\mathbf{z}_i \sim \mathcal{N}(0, \sigma^2)$, and $\mathbf{u}'_i \sim \text{unif}(\mathcal{V}_{1,i}^Q)$. We can rewrite the above expression in the form

$$\mathbb{P}\left(\|\sum_{j \in \Omega} h_{ij}(\mathbf{x}_j^{(k_j)} - \mathbf{x}_j^{(k'_j)}) + \mathbf{z}_i - \mathbf{u}'_i\|^2 \leq n\,\sigma_c^2, \forall i \in \Omega^c\right).$$

with the understanding that $h_{ij}$ is only non zero if $i \in \mathcal{M}_l$ and $j \in \mathcal{M}_{l-1}$ for some $l = 1, \ldots, l_d$. Note that we can immediately apply Lemma 7.3 by identifying $\mathcal{S}_i$ in the lemma as $\mathcal{B}(\sqrt{n\,\sigma_c^2})$, and switch from the discrete distribution over the lattice points $p_r^{-1}\Lambda^T \cap \mathcal{V}^T$ for $\mathbf{x}_j^{(k_j)}$ and $\mathbf{x}_j^{(k'_j)}$, $j \in \Omega$ to a Gaussian distribution. More precisely, the above probability is upper bounded by

$$\left((1 + \epsilon_4(\Lambda^T))^{N_R} e^{n\epsilon_1(\Lambda^T)+n\epsilon_2}\right)^{2N_L} \left(e^{n\epsilon_2(\Lambda_{1,i}^Q)}\right)^{N_R} \mathbb{P}\left(\|\sum_{j \in \Omega} h_{ij}(\tilde{\mathbf{x}}_j - \tilde{\mathbf{x}}'_j) + \tilde{\mathbf{z}}_i\|^2 \leq n\,\sigma_c^2, \forall i \in \Omega^c\right), \quad (72)$$

where $\tilde{\mathbf{x}}_j, \tilde{\mathbf{x}}'_j, j \in \Omega$ are independent $\sim \mathcal{N}(0, \sigma_x^2 I_n)$ with

$$\sigma_x^2 = (1 + p_r^{-1})^2 (\rho_{cov}(\Lambda^T))^2 \frac{G_n^*}{G(\Lambda^T)} \sigma^2(\Lambda^T),$$

and $\tilde{\mathbf{z}}_i, i \in \Omega$ are independent $\sim \mathcal{N}(0, \sigma_z^2 I_n)$,

$$\sigma_z^2 = (1 + \epsilon_5)^{2N_L}\left(1 + \frac{G_n^*}{G(\Lambda_{1,i}^Q)}(\rho_{cov}(\Lambda_{1,i}^Q))^2 \sigma^2(\Lambda_{1,i}^Q)\right)$$

where all $\epsilon_1(\Lambda^T)$, $\epsilon_2$, $\epsilon_2(\Lambda_{1,i}^Q)$, $\epsilon_4(\Lambda^T)$, $\epsilon_5 \to 0$ as $n \to \infty$. Furthermore $\sigma_x^2 \to P$ and $\sigma_z^2 \to 2$ as $n \to \infty$ since all $\Lambda^T$, $\Lambda^Q$ and $\Lambda_{1,i}^Q$ are Rogers-good.

The probability in (72) can be upper bounded as follows:

$$\mathbb{P}\left(\|\sum_{j \in \Omega} h_{ij}(\tilde{\mathbf{x}}_j - \tilde{\mathbf{x}}'_j) + \tilde{\mathbf{z}}_i\|^2 \leq n\,\sigma_c^2, \forall i \in \Omega^c\right)$$

$$\leq \mathbb{P}\left(\|H(\tilde{X} - \tilde{X}') + \tilde{Z}\|_2^2 \leq N_R\,n\,\sigma_c^2\right) \quad (73)$$

$$= \mathbb{P}\left(\|\Sigma(\tilde{X} - \tilde{X}') + \tilde{Z}\|_2^2 \leq N_R\,n\,\sigma_c^2\right) \quad (74)$$

$$\leq \mathbb{P}\left(\sum_{i=1}^{\min(N_R, N_L)} \|\sigma_i(\tilde{\mathbf{x}}_i - \tilde{\mathbf{x}}'_i) + \tilde{\mathbf{z}}_i\|^2 + \sum_{i=1}^{(N_R-N_L)^+} \|\tilde{\mathbf{z}}_i\|^2 \leq N_R\,n\,\sigma_c^2\right), \quad (75)$$

where $H$ is the $N_R \times N_L$ transfer matrix from the nodes in $\Omega$ to the nodes in $\Omega^c$ and $\Sigma$ is a diagonal matrix containing the singular values $\sigma_i, i = 1, \ldots, \min(N_R, N_L)$ of $H$. $\tilde{X}$ and $\tilde{X}'$ are $N_L \times n$ matrices, their $j$'th row containing the vectors $\tilde{\mathbf{x}}_j$ and $\tilde{\mathbf{x}}'_j$ respectively. $\tilde{Z}$ is $N_R \times n$ matrix, its $i$'th row containing the vector $\tilde{\mathbf{z}}_i$. The entries of the matrix $\tilde{X} - \tilde{X}'$ are i.i.d. with distribution $\mathcal{N}(0, 2\sigma_x^2)$ and the entries of the matrix $\tilde{Z}$ are i.i.d. with distribution $\mathcal{N}(0, \sigma_z^2)$. Inequality (73) follows from the definition of the Frobenius norm for matrices. (74) is obtained by replacing $H$ with its singular value decomposition $U\Sigma V^\dagger$ and noting that for any matrix $A$, $\|U^\dagger A\|_2 = \|A\|_2$ when $U$ is unitary. Moreover, the distribution of $U^\dagger \tilde{Z}$ is the same as $\tilde{Z}$ and the distribution of $V^\dagger(\tilde{X} - \tilde{X}')$ is the same as $(\tilde{X} - \tilde{X}')$.

The probability in (75) can be bounded using Lemma 7.2. For any $t > 0$,

$$\mathbb{P}\left(\sum_{i=1}^{\min(N_R,N_L)} \|\sigma_i(\tilde{\mathbf{x}}_i - \tilde{\mathbf{x}}'_i) + \tilde{\mathbf{z}}_i\|^2 + \sum_{i=1}^{(N_R-N_L)^+} \|\tilde{\mathbf{z}}_i\|^2 \leq N_R\, n\, \sigma_c^2\right) \leq$$
$$e^{-\frac{n}{2}\left(\sum_{i=1}^{\min(N_R,N_L)} \log\left(1+2(2\sigma_i^2\sigma_x^2+\sigma_z^2)t\right) + \sum_{i=1}^{(N_R-N_L)^+} \log\left(1+2\sigma_z^2 t\right) - 2tN_R\,\sigma_c^2\right)}.$$

Choosing $t = 1/2\sigma_c^2$, yields an exponent

$$-\frac{n}{2}\left(\sum_{i=1}^{\min(N_R,N_L)} \log\left(1 + \frac{2\sigma_i^2\sigma_x^2 + \sigma_z^2}{\sigma_c^2}\right) + \sum_{i=1}^{(N_R-N_L)^+} \log\left(1 + \frac{\sigma_z^2}{\sigma_c^2}\right) - N_R\right)$$

in the above expression. We have

$$\frac{2\sigma_i^2\sigma_x^2 + \sigma_z^2}{\sigma_c^2} \to \frac{\sigma_i^2 P + 1}{(1+\epsilon)}, \qquad \frac{\sigma_z^2}{\sigma_c^2} \to \frac{1}{1+\epsilon},$$

as $n \to \infty$. Combining everything together yields,

$$\mathbb{P}\left(\|\sum_{j\in\Omega_{l-1}} h_{ij}(\mathbf{x}_j^{(k_j)} - \mathbf{x}_j^{(k'_j)}) + \mathbf{z}_i - \mathbf{u}'_i\|^2 \leq n\,\sigma_c^2, \forall i \in \Omega^c\right)$$
$$\leq e^{-\frac{n}{2}\left(\sum_{i=1}^{\min(N_R,N_L)} \log\left(1+\sigma_i^2 P\right) - N_R(1+\log(1+\epsilon)) + o_n(1)\right)}.$$

In the last expression we identify $\frac{1}{2}\sum_{i=1}^{\min(N_R,N_L)} \log\left(1 + \sigma_i^2 P\right)$ as $I(X_\Omega; HX_\Omega + Z_{\Omega^c})$, where $X_\Omega$ is an $N_L \times 1$ Gaussian vector with i.i.d entries of variance $P$ and $Z_{\Omega^c}$ is an $N_R \times 1$ Gaussian vector with i.i.d entries of variance $\sigma^2$ and $H$ is the corresponding transfer matrix between nodes in $\Omega$ and $\Omega^c$.

*Proof of Lemma 5.3:* Note that a priori the random variables $\tilde{\mathbf{y}}_i^{(k'_i)}, i \in \Omega$ in (39) for a fixed set of indices $\{k'_i\}_{i\in\Omega}$ are independent and uniformly distributed over $\mathcal{V}^Q$. This is because the quantization codebook at each relay is chosen at random from the ensemble of Section III (note that the construction of the ensemble induces a uniform mapping between the indices $k'_i = 1, \ldots, p_r^{k_r}$ and the corresponding lattice points) and $\tilde{\mathbf{y}}_i^{(k'_i)}$ is obtained by dithering $\hat{\mathbf{y}}_i^{(k'_i)}$ over the Voronoi region $\mathcal{V}_{1,i}^Q$ in (23). As a result, $\tilde{\mathbf{y}}_i^{(k'_i)}$ for $i \in \Omega$ are independent continuous random variables uniformly distributed over $\mathcal{V}^Q$. Moreover, this is still the case conditioned on the events

$$\mathcal{A}_i = \{\|(\sum_{j\in\Omega_{l-1}} h_{ij}(\mathbf{x}_j^{(k_j)} - \mathbf{x}_j^{(k'_j)}) + \mathbf{z}_i - \mathbf{u}'_i) \mod \Lambda^Q\|^2 \leq n\,\sigma_c^2\}, \qquad i \in \Omega^c$$

Note that the event in the conditioning governs the set of random variables $\{\mathbf{x}_i^{(k_i)}, \mathbf{x}_i^{(k'_i)}, i \in \Omega\}, \{\mathbf{z}_i, \mathbf{u}_i, i \in \Omega^c\}$. $\tilde{\mathbf{y}}_i^{(k'_i)}, i \in \Omega$ are independent from these random variables, therefore conditioned on $\mathcal{A}_i, i \in \Omega^c$, $\tilde{\mathbf{y}}_i^{(k'_i)}, i \in \Omega$ are still independent uniformly distributed over $\mathcal{V}^Q$. By the Crypto Lemma, the random variables

$$\nu_i = \tilde{\mathbf{y}}_i^{(k'_i)} - \sum_{j\in\Omega_{l-1}^c} h_{ij}\mathbf{x}_j^{(k_j)} - \sum_{j\in\Omega_{l-1}} h_{ij}\mathbf{x}_j^{(k'_j)} \mod \Lambda^Q, \qquad i \in \Omega^c$$

are also uniformly distributed over $\mathcal{V}^Q$ and is independent of

$$\sum_{j\in\Omega_{l-1}^c} h_{ij}\mathbf{x}_j^{(k_j)} + \sum_{j\in\Omega_{l-1}} h_{ij}\mathbf{x}_j^{(k'_j)}.$$

This is due to the fact that $\tilde{\mathbf{y}}_i^{(k_i')}$ is independent of this term. Therefore (46) is upper bounded by

$$\sum_{\substack{k_i', i \in \Omega \\ k_i' \neq k_i}} \mathbb{P}\left(\mathcal{B}_i, i \in \Omega \,|\, \mathcal{A}_i, i \in \Omega^c\right) = e^{|\mathcal{N}_\Omega| n R_r} \prod_{i \in \mathcal{N}_\Omega} \mathbb{P}\left(\|\nu_i\|^2 \leq n \sigma_c^2\right)$$

$$\leq e^{\frac{1}{2} n |\mathcal{N}_\Omega| \log \sigma^2(\Lambda^Q)} e^{-\frac{1}{2} n |\mathcal{N}_\Omega| \left(\log\left(\frac{\sigma^2(\Lambda^Q)}{\sigma_c^2}\right) - 1 + \frac{\sigma_c^2}{\sigma^2(\Lambda^Q)} - o_n(1)\right)}$$

$$\leq e^{\frac{1}{2} n |\mathcal{N}_\Omega| n (\log 2(1+\epsilon) + 1 + o_n(1))}$$

where used Lemma 5.4 and the fact that $R_r = \frac{1}{2} \log \sigma^2(\Lambda^Q)$ from (19). □